%% file: arxiv.tex
\documentclass{sig-alt}
\usepackage{txfonts} 

\usepackage[numbers]{natbib}
\setcitestyle{square,citesep={;},aysep={},yysep={;}}
\usepackage{url}
\usepackage{paralist}
\usepackage{color}

\usepackage{amssymb}
\usepackage{graphics}
\usepackage{comment}
\usepackage{algpseudocode}
\usepackage{algorithm}
\usepackage{booktabs}
\usepackage{verbatim}
\usepackage{etoolbox}
\usepackage{nth}
\usepackage[smallbib]{fixacm}
\usepackage{xspace}
\input{macro} 

\patchcmd{\maketitle}{\@copyrightspace}{}{}{}

\begin{document}

\title{Concise Summarization of Heterogeneous Treatment Effect Using Total Variation Regularized Regression}

\numberofauthors{5} 
\author{
\alignauthor
Alex Deng\thanks{First two authors contributed equally to this work.}\\
       \affaddr{Microsoft}\\
       \email{alexdeng@microsoft.com}
\alignauthor
Pengchuan Zhang\textsuperscript{*}\thanks{Work done as an intern in Microsoft Analysis and Experimentation.}\\
       \affaddr{California Institute of Technology}\\
       \email{pzzhang@cms.caltech.edu}
\alignauthor
Shouyuan Chen\\
\affaddr{Microsoft}\\
       \email{shouch@microsoft.com}
\and
\alignauthor
Dong Woo Kim\\
\affaddr{Microsoft}\\
       \email{dok@microsoft.com}
\alignauthor
Jiannan Lu\\
\affaddr{Microsoft}\\
       \email{jiannl@microsoft.com}
}
\date{August, 2016}

\maketitle
\begin{abstract}
Randomized controlled experiment has long been accepted as the golden standard for establishing causal link and estimating causal effect in various scientific fields. Average treatment effect is often used to summarize the effect estimation, even though treatment effects are commonly believed to be varying among individuals. In the recent decade with the availability of ``big data'', more and more experiments have large sample size and increasingly rich side information that enable and require experimenters to discover and understand heterogeneous treatment effect (HTE). There are two aspects in HTE understanding, one is to predict the effect conditioned on a given set of side information or a given individual, the other is to interpret the HTE structure and summarize it in a memorable way. The former aspect can be treated as a regression problem, and the latter aspect focuses on concise summarization and interpretation. In this paper we propose a method that can achieve both at the same time. This method can be formulated as a convex optimization problem, for which we provide stable and scalable implementation.
\end{abstract}

\section{Introduction}
\label{sec:introduction}
A large body of research across many disciplines seeks to draw inferences from data about the causal effect of a treatment. Examples include medical studies about effects of drugs on health outcomes \citep{bartroff2012sequential,tsiatiscovariateadj}, studies of impacts of advertising or marketing offers on consumer purchases \citep{lewis2011here,chan2010evaluating}, evaluations of the effectiveness of government programs or public policies \citep{heckman1991randomization,athey2016econometrics}, and more recently in online experimentation under the name ``A/B tests" (large-scale online randomized experiments), used widely by web facing technology firms \citep{Kohavi:2009, Kohavi:2013, googlesurvey, bakshy2014designing, gomez2016netflix}. 

Historically, many works on digital experimentation have been focused on trustworthy analysis of the average treatment effect (ATE), where trustworthiness includes unbiased identification of ATE as well as correct assessment of its uncertainty (confidence interval, Type-I error and False Discovery Rate, sequential test, etc.) \citep{Kohavi:2013, DengBayesAB, johari2015always}. It is commonly understood among researchers and practitioners that individual treatment effects may differ in magnitude and even have opposite direction. This is called heterogeneous treatment effect (HTE) \citep{athey2015recursive, taddy2016nonparametric,tian2014simple}. There are two distinct reasons why understanding HTE is important.
\begin{compactenum}
   \item Personalized treatment: When there are many different potential treatments, e.g. policies, drug dosages, etc., we want to find the best treatment that maximize a pre-specified metric or utility for each individual. To achieve that, we will need to be able to predict treatment effect for every candidate treatment on every potential individual.
   \item Improvement and agile iteration: For a promising treatment with positive average effect, we want to find areas where the treatment can be further improved. One obvious direction is to identify scenarios where the treatment performs relatively bad, or even shows negative effects. Understanding HTE leads to sub-populations representing areas-to-improve.  
\end{compactenum}
The first point put more emphasis on prediction and can be treated as a kind of supervised learning problem \citep{tian2014simple, athey2015recursive, wager2015estimation, taddy2016nonparametric}. The second point focus heavily on interpretation and summarization. The structure of HTE, i.e. how treatment effects differ between individuals, is of great interest as compared to only the value of individual treatment effect. In fact, even for the purpose of personalized treatment, it is also desired to have a interpretable and memorable result \citep{athey2016econometrics}.

In this paper, we aim at analyzing the heterogeneity of treatment effects in A/B tests, where the understanding of HTE structure is more important. This is because the goal of A/B tests is to improve a software, server side or client side. Code maintenance can be expenseive if we were to keep different branches of the same code just to achieve personalization. In reality, it is much easier to improve one version and make it work better for all users than maintaining ensembles of code. In addition to carving out the structure of HTE using an interpretable model, we also strive for a \emph{concise summarization} so it is memorable and actionable for product owners and engineers. Because there are many ways of defining sub-populations, we adopt the following definitions through out the paper.
\begin{defi-no}
A \textbf{covariate} is an observed attribute that can be used to split the population into sub-populations, e.g. browser, device, location, etc. A covariate can be continuous or discrete. A discrete covariate consists of a number of \textbf{levels}, e.g. different browser types. The number of levels for a covariate is called its \textbf{size}. A discrete covariate is often called a \textbf{segment type} with each of its levels called a \textbf{segment}. Discrete covariate can be either ordered, e.g. date, income levels, or categorical without ordering. 
\end{defi-no}

We provide two concrete problems to illustrate the type of HTE we care about. The first one is a typical application in A/B testing and the second one is an extension outside of controlled experiments. 
\begin{problem}\label{pb:1}
Suppose a treatment showed positive treatment effect on all browsers except several old versions of Internet Explorer for a period of two weeks. In addition, during the second week the treatment effects for Android devices dipped by a large amount. In this case the HTE can be decomposed into two pieces of information. First, this treatment does not work well for old IE browsers, possibly due to lack of legacy browser support. Second, something changed between the first and the second week for Android devices. Based on these two pieces of information, engineers can then prioritize work items to improve the legacy browser support and also works with product analysts to understand whey Android devices behaves different for the second week. 
\end{problem}

\begin{problem}\label{pb:2}
We are monitoring a vital KPI metric of a business and observed an unexpected drop between this week and previous week. Since there is no sudden change in our product, we believe the drop is not systematically across the board, but rather due to some small sub-population, e.g. a set of mobile devices, geo-locations, etc. The task is to figure out these sub-populations that explain majority of the change so we can zoom in and fix the problem. 
\end{problem}
The second example is a diagnosis problem. Although the pre-post comparison is not a controlled experiment and the difference between the two periods could be due to differences of many confounding factors, we are interested in identifying subgroups that defines the structure of heterogeneous pre-post differences. The problem of HTE structure and heterogeneous pre-post difference structure are essentially the same.  

A common characteristic of the two problems above is that in both cases we want to summarize HTE structure in a very concise way. The obvious challenge is that the dimension of covariates is large and they can be further combined together to create higher order population segmentation. An ideal result is to identify lower order effects, i.e. sub-populations defined by small number of covariates. This is because product owners can only afford deeper investigations into a small number of cases. To make our result actionable, we have to bet in Sparsity-of-effects principle \citep{box2005statistics} since results with too complicated HTE structure will confuse people and discourage action taking. 

This paper makes the following contributions:
\begin{compactenum}
   \item We propose using total-variation regularized regression (TV Regression) to produce concise summarization of HTE and the diagnosis problem. Our method can be applied to all discrete covariates for both categorical and ordered. We describe and implement an ADMM algorithm to efficiently solve the convex optimization problem. 
   \item When sizes of covariates differ a lot, we allow penalizing total variation of different covariates differently. We provide a pre-tuning algorithm to help us find the best weights and our algorithm naturally generalizes to other problems such as Grouped Lasso \citep{yuan2006model}. We propose a post-screening stage that can be optionally added and further improve the result of TV Regression.
   \item We applied TV regression to both synthetic data and real experiment data. We also demonstrate the necessity to use weight pre-tuning. 
\end{compactenum}

The rest of the paper is structured as follows. We first describe the background and existing approaches in literature to motivate our method. In Section~\ref{sec:TVregularized} we introduce Total-Variation regularized regression to achieve simultaneous HTE prediction and concise summarization. We demonstrate the effectiveness of TV regression using both synthetic and real experiment data in Section~\ref{sec:examples}, as well as the importance of adaptive between-graph weights. We also included a real application of TV regression for a diagnosis problem.   

\section{Background}\label{sec:setting}
Suppose we have $N$ \iid data points labelled $i = 1, 2, \dots, N$, each of which consists of a covariate vector $X^i = (x^i_1,x^i_2,\dots,x^i_D)\in \R^D$, a response $Y_i\in \R$ and a treatment indicator $W_i \in \{0, 1\}$. These covariates can be categorical, e.g. gender, market, and can also be ordered, such as user age group, user visit frequency, date and time, etc. We do not consider continuous covariates and assume they can always be discretized into ordered covariates. This will not affect the generality of our approach as discussed later. Following the potential outcomes framework \citep{Imbens:2015, Rosenbaum1983, Rubin:1974}, we posit the existence of potential outcomes $Y_i^{(1)}$ and $Y_i^{(0)}$ corresponding respectively to the response the $i$-th user would have experienced with and without the treatment, and define the treatment effect at $x$ as
\begin{equation*}
	\tau(x) = \Expect[Y^{(1)} - Y^{(0)} | X = x],
\end{equation*}
where $(X,Y^{(1)}, Y^{(0)})$ has the same distribution of $(X^i,Y_i^{(1)}, Y_i^{(0)}),i=1,\dots,N$. Our goal is to give an estimation of $\tau(x)$, which is the regression of $Y^{(1)} - Y^{(0)}$ with covariate vector $X$. 

This is a regression problem if $Y_i^{(1)} - Y_i^{(0)}$ are directly observed. However, we can only ever observe one of the two potential outcomes $Y_i^{(1)}$ and $Y_i^{(0)}$ for $i^{\text{th}}$ data point. Let the observed outcome be $$Y_i^{\textrm{obs}} = W_i Y_i^{(1)} + (1-W_i)Y_i^{(0)}. $$ There are in general two different views to deal with this problem in the literature. 

The first approach is to model the counter-factual pair $(Y^{(1)}_i, Y^{(0)}_i)$. The classic parametric model is a linear model with interaction terms between treatment assignment indicator $W$ and covariates $X$: 
\begin{align}\label{eq:lm}
    Y^{\textrm{obs}}_i = f(X^i) + W_i \times \left ( \mu + \beta^T X^i \right ) + \epsilon_i,
\end{align}
where $f(X^i)$ is usually also a linear function, $\mu$ is the constant effect and $\beta \in \R^D$ is the coefficient vector that encodes the HTE as a linear function of $X^i$. When dimensionality is high, a $\ell_1$ regularization \citep{tibshirani1996regression} is typically employed to enforce sparsity and prevent over-fitting. This approach has a long history and has appeared in different variations in the literature \citep{tian2014simple,imai2013estimating,taddy2016nonparametric}. This approach also includes tree based method where two tree based models are fit to for regression of $Y^{(1)}$ and $Y^{(0)}$ respectively.  

A second approach is to directly estimate $\tau(x)$. Given a covariate vector $X = x$, if we can find enough data points with exactly the same $x$ in the experiment, and because treatment assignment is completely random, a unbiased estimator of $\tau(x)$ is just
\begin{equation}\label{def:tauhat}
	\hat{\tau}(x) = \bar{y}^{(1)}(x) - \bar{y}^{(0)}(x),
\end{equation}
where $\bar{y}^{(1)}(x)$ and $\bar{y}^{(0)}(x)$ is the treatment average and control average on the subset of data points having the same covareate $x$:
\begin{equation*}
	\overline{y}^{(w)}(x) = \frac{\sum_{X^i = x, W_i = w} Y_i}{N^{(w)}(x)}, \quad N^{(w)}(x) := \sum_{X^i = x, W_i = w} 1.
\end{equation*}
This may not be effective when dimensionality $D$ is large and $N^{(w)}(x)$ may be small for some $x$. But in reality, under the assumption of Sparsity-of-effects \citep{box2005statistics}, HTE might only depends on only a few covariates in the vector $x$, we can imagine a neighborhood of $x$ will have the same $\tau(x)$. This leads to Nearest Neighbor based method and nonparametric smoothing of \eqref{def:tauhat}. \citet{athey2015recursive} and \citet{wager2015estimation} used decision tree based method to adaptively partition the covariate space to achieve simultaneous neighborhood matching (leaves in tree) and HTE estimation. Different splitting criteria were used to achieve slightly different goals. For example, \citep{athey2015recursive} suggested to build a CART like decision tree by maximizing between leaf variance so the result tree can be used for both prediction and also interpretable. 

\begin{figure*}[!tb]
    \centering
    \includegraphics[width=0.95\textwidth,height=0.3\textwidth]{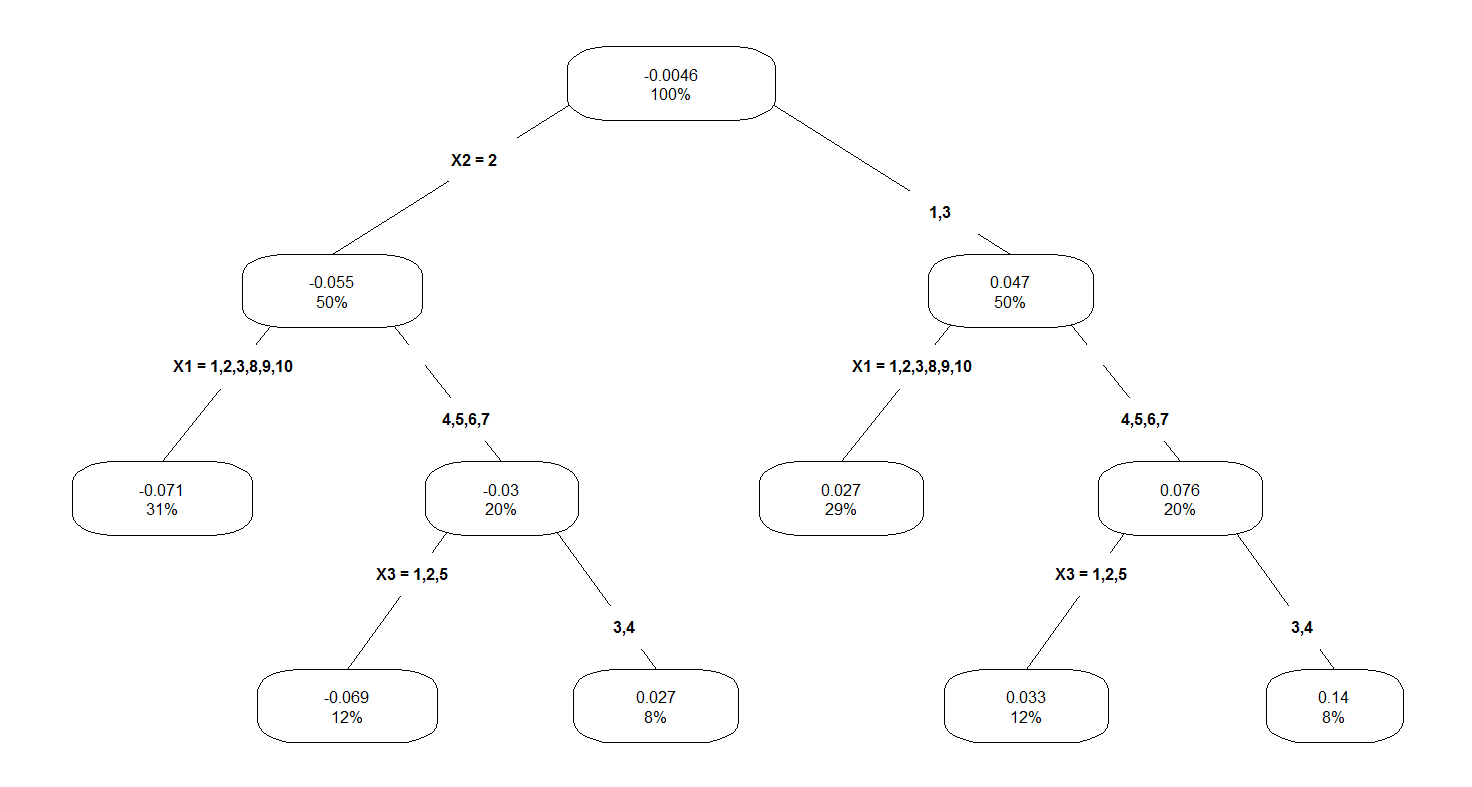}
    \caption{Causal Tree for Example~\ref{ex:1}. Number in each leaf is the predicted treatment effect and percentages are proportions of the whole population. The decision tree fitted the HTE quite well, but hard to summarize and be memorized.}
    \label{fig:ctree}
\end{figure*}

\subsection{Issues of Existing Approaches}\label{subsec:issue}
For applications like Problem~\ref{pb:1} and \ref{pb:2}, we found none of the currently existing approaches provides satisfying results. Consider the following example.
\begin{ex}\label{ex:1}
There are 4 categorical covariates $X_1,\dots, X_4$. $X_1$ has 10 levels, each 10\%. $X_2$ has 3 levels, with proportion 20\%, 50\% and 30\%. $X_3$ have 5 equally proportion levels and $X_4$ have 4 levels with 10\%,  40\%, 30\%, 20\%. Let $X_i$$:$$j$ be the jth level of $X_i$. We then put a simple ground truth effect as this: $X_2$$:$$2$ has first order effect of -0.1; there exist a second order effect of 0.1 on $X_1$$:$$(4,5,6,7)$ $\cap$ $X_3$$:$$(3,4)$. In addition, there is a constant global effect 0.03. $10,000$ \textit{i.i.d.} treatment and control observations are simulated from $N(0, 0.1)$ distribution respectively with treatment effect added to treatment observations according to the prescribed design.
\end{ex}

The ground truth HTE structure for Example~\ref{ex:1} can be concisely summarized by two numbers: one first order effect and one second order effect (global effect is not of interest). Note that the second order effect is defined as a block structure. In practice, block structure are common building blocks for HTE because it stands for clustering. These clusters often have semantic meaning. For example, web browsers are naturally classified by rendering engines and versions, similarly mobile devices are clustered by brand and operational systems. Block-wise structure is also very easy to memorize and allow product owners to clearly define where to further investigate. Although in Example~\ref{ex:1} the first order effect is ``point-wise'', i.e. consist of one level only. In general they can also be clusters of different levels(segments) for each covariate(segment type). 

The setup of lower order effects suggests classic linear model with $\ell_1$ regularization will work well because the model captures first and second order effects as its coefficients in the linear regression. For an appropriated tuned parameter, we showed that Lasso does recover all first order and second order effects. However it suffers from two issues. First it could not cluster block-wise second order effects together. Instead, it will output 8 different coefficients for all combinations of $X_1$:$(4,5,6,7)$ and $X_3$:$(3,4)$. Second, we found Lasso still tend to output false positives, i.e. non-zero coefficients which are known to be 0 in the ground truth. We provide more details in Section~\ref{subsec:comparelasso} and similar results were shown in \citep{tian2014simple}. The main issue of Lasso failing to discover block structures is expected because the model does not perform clustering or look for block structure. Tree based method, however, naturally produce block-wise constant prediction. We applied Causal Tree \citep{athey2015recursive} (without honest splitting) and the result is shown in Figure~\ref{fig:ctree}. We point out at least two issues with tree based method:
\begin{compactenum}
   \item Although single decision tree is interpretable. It is not concise. This is because it has fixed number of branches in each split, usually binary. It is good at showing higher order effect as each leaf is block-wise, but it does not summarize lower order effect well. To convince yourself, look at Figure~\ref{fig:ctree} and imagine how one can recover the simple ground truth structure represented by 3 lower order effects using this tree. Note that the prediction from this tree is good, just hard to summarize.
   \item Tree relies on pruning to avoid over-fitting. But pruning only controls tree size without prevent over-fitting in earlier stage of splittings. This problem can be alleviated by putting aside a test set and use test data performance to optimize splitting on training data. This is the idea behind honest splitting in \citep{athey2015recursive, wager2015estimation}. Others suggested using statistical testing based splitting criteria to avoid over-fitting in each splitting \citep{hothorn2006unbiased}. These modifications further complicate the procedure making it even harder to be used in practice on big data. 
\end{compactenum}

To overcome shortcomings of both approaches, we propose a new approach. We chose to use an additive model, similar to Lasso to summarize HTE as a set of lower order effects. But instead of fitting this additive model with sparsity requirement for nonzero coefficients, we also want to simultaneously recover block-wise structures among non-zero coefficients. This is known as \emph{simultaneous clustering and regression} \citep{bondell2008simultaneous}. 

\section{Simultaneous clustering and \newline regression by Total-Variation \newline regularization}
\label{sec:TVregularized}
The idea of simultaneous clustering and regression naturally follows the discussions in Section~\ref{subsec:issue}. We use an additive model with $\ell_1$ penalty on nonzero coefficients, and want those nonzero coefficients to form block-wise clusters so that we can easily summarize the structure. In both applications of HTE and diagnosis problem, we limit the type of covariates as discrete. This is because nonlinear effect of a continuous covariate is hard to define in an additive model but can easily be approximated by linear effect of a discretized and ordered covariate. When all covariates are discrete, an additive model can be written as:  
\begin{align}\label{eqn:additive}
	\tau(x) = u_0 + \sum_{k=1}^K u_k(x)
\end{align}
where $u_0$ represents the constant global treatment effect and $\{u_k(x)\}_{k=1}^K$ are different kinds of heterogeneous effects. The constant term $u_0$ is not of much interest when studying the heterogeneous treatment effect. Additive components can be defined with one covariate (first order effect) or multiple covariates (higher order effects). A model with up to second order effects is written as:
\begin{equation}\label{eqn:secondordermodel}
    \tau(X_i) = u_0 + \sum_{d=1}^D u_d(x_d) + \sum_{1 \le d < f \le D} u_{d,f}(x_d, x_f),
\end{equation}
where $D$ is the number of covariates.

Given the additive model of $\tau(x)$, we can fit the model to predict $Y^{\textrm{obs}}$ as in \eqref{eq:lm}. When viewing the problem as predicting individual $Y^{\textrm{obs}}$s, we need to deal with a very big dataset, quite often millions of $Y$'s for a large scale online services. On the other hand, for any combination of covariates $X^i = x$, we observe sufficient statistics
$$
(\bar{y}^{(1)}(x), \bar{y}^{(0)}(x), \hat{\tau}(x), N^{(1)}(x), N^{(0)}(x)).
$$
Turns out, when we treat $y^{(0)}(x) = \Expect[Y_i^{(0)}| X^i = x]$ as model free, minimizing squared error for all $Y^{\textrm{obs}}$ (divided by their variances) is equivalent to the following optimization problem\footnote{Derivation is straightforward but tedious. See Appendix~\ref{app:equiv}.}: 
\begin{equation}\label{eqn:master}
    \min_{\tau} \frac{1}{2}\sum_{x} M_{e}(x) (\hat{\tau}(x) - \tau(x))^2,
\end{equation}
where 
\begin{equation}\label{def:M}
    M_{e}(x) = \frac{N^{(0)}(x)N^{(1)}(x)}{N^{(0)}(x)\Var^{(1)}(x)+N^{(1)}(x)\Var^{(0)}(x)}
\end{equation}
can be viewed as the effective sample size for the noisy measurement $\hat{\tau}(x)$ of the treatment effect at covariate combination $x$ and $\Var^{(w)}(x)$ are conditional variances of $Y^{(w)}$ at $x$.

Solving \eqref{eqn:master} can be seen as trying to predict $\tau(x)$ directly, as in \citep{athey2015recursive,wager2015estimation}. For moderate number of covariates or small size covariates, the number of covariate combinations is much smaller than the size of the original data, e.g. 10 binary covariates generate only 1k combinations, thus reducing the complexity of the optimization problem. When the number of covariate combinations is large, we apply a pre-screening stage to prune out majority of irrelevant covariates. More details in Section~\ref{subsec:prescreen}. 
To cluster these effects together, we add penalty $|u_k(x) - u_{k}(x')|$ for every pair of $(x, x')$ that we would like to cluster together if their effects are close. The latter $\ell_1$ penalty is called total variation regularization for which we give detailed definition in the next section. TV regression solves the following convex optimization problem:
\begin{align}
\min_{\tau} \quad & \frac{1}{2}\sum_{x} M_{e}(x) (\hat{\tau}(x) - \tau(x))^2 + \lambda TV(\tau) \nonumber \\
\text{s.t.}\quad &\tau(x) = u_0 + \sum_{k=1}^K u_k(x), \label{eqn:tvregreg}
\end{align}
where $TV(\tau)$ is a given total variation penalty and $\lambda > 0$ is a parameter balancing the approximating accuracy and the regularization. We didn't add $\sum_{k,x} |u_k(x)|$ in \eqref{eqn:tvregreg} because we can easily include these terms within $TV(\tau)$.

\subsection{Total Variation on Graphs}
\label{subsec:tvongraphs}
Total variation penalty was introduced in images denoising \cite{rudin1992nonlinear, meyer2001oscillating, osher2003image}, where it is well-known to outperform $\ell_2$-regularized estimators like Laplacian smoothing (and also Laplacian eigenmaps) for its ability to better preserve sharp edges and object boundaries. In our setting, we can view each covariates combination $x$ as a pixel in a high dimensional image with global effect as background image and lower order effects as the foreground. In statistics literature, similar ideas are called Fused Lasso, trend filtering and Generalized Lasso \citep{tibshirani2005sparsity, tibshirani2011solution, wang2015trend, petersen2015fused, petersen2015fused, petersen2016convex}. In most cases, covariates are assumed to be ordered and the $\ell_1$ penalties of total variation are applied to pairs of neighborhood points/pixels. In our applications, a lot of covariates are categorical without ordered topology. To unify the two cases here we define TV penalty on a graph.

Let $G = (V, E)$ be a graph with nodes labelled $V = \{1, 2, \dots, n\}$ and edges $E = \{e_1, e_2, \dots, e_m\}$. Let $u : V \to \R$ be a function defined on $G$. The total variation of $u$ is defined as 
\begin{equation}\label{def:tvgraph0}
    TV(u) = \sum_{e \in E} |u(e_1) - u(e_0)|,
\end{equation}
where $e_0$ and $e_1$ are the two vertices of an edge $e$.

Every covariate $x_d$ defines a graph, denoted as $G_d = (V_d, E_d)$, where vertices $V_d$ are all its levels. These graphs can either be fully connected or encoding topological information. For example, for covariates with order, we connect all the consecutive nodes; for weekdays or months, we also connect Sunday to Monday and December to January to make loops; for combinatorial covariates, we can simply have a complete graph or use extra domain knowledge. For second order effect involving two covariates, its graph can be defined by the tensor product of the two single covariate graph, i.e., $G_d\times G_f = (V_d\times V_f, E_d\times E_f)$ where the vertex set is the Cartesian product of $V_d\times V_f$ and edge set $E_d\times E_f$ is defined such that there is an edge between $(x_d, x_f)$ and $(x'_d,x'_f)$ if and only if $x_d\sim x'_d$ in $E_d$ and $x_f=x'_f$ or $x_d = x'_d$ and $x_f \sim x'_f$ in $E_f$. Higher order graph is defined similarly.  

For any term $u_k(x)$ in the additive model \eqref{eqn:additive}, a naive choice is to define TV by Eqn.~\eqref{def:tvgraph0} for each term and then sum them up to define $TV(\tau)$. However, in many applications, the sizes of different covariates vary a lot. For example, gender has only 2 states, but product category has over hundreds. Simply putting equal weights to every edge as in~\eqref{def:tvgraph0} does not work well due to this heterogeneity. We define
\begin{equation}\label{eqn:tvforadditive}
	TV(\tau) = \sum_{k=1}^K w_k TV(u_k),
\end{equation}
in which $\{w_k\}_{k=1}^K$ are some to-be-determined weights. A first-order effect $u_d(x_d)$ has the following modified TV
\begin{equation}\label{eqn:tvfor1st}
	TV(u_d) = (1-\alpha)\sum_{e \in E_d} |u_d(e_1) - u_d(e_0)| + \alpha \sum_{v\in V_d} |u_d(v)|
\end{equation}
and a second-order effect $u_{d, f}(x_{d}, x_{f})$ has TV defined by
\begin{align}
	TV(u_{d, f}) & = (1-\alpha)\sum_{e \in E_d\times E_f} |u_{d,f}(e_1) - u_{d,f}(e_0)|  \nonumber \\
    &	+ \alpha \sum_{v\in V_d\times V_f} |u_{d,f}(v)|. \label{eqn:tvforhighorder}
\end{align}
Note that we've incorporated Lasso penalty $|u_k(x)|$ into TV penalty so we no longer need it in \eqref{eqn:tvregreg}. TV for higher order effects can be defined similarly, but we believe that first and second order effects are most useful for practice. 

Our definition of TV on graphs has the following properties:
\begin{compactitem}
\item We define the TV separately for different terms because each term so we encourage effect clustering within each additive term. This separation also makes it computationally more efficient. 
\item In Section~\ref{subsec:weights} we introduce a simple procedure for choosing weights $\{w_k\}_{k=1}^K$ appropriately to deal with the issue of different covariate sizes. Thus we do not need to treat these extra weights as tuning parameters. 
\item We can also adjust the weights between TV and $\ell_1$ penalty by adjusting $\alpha \in [0,1]$. When $\alpha = 1$ TV regression reduces to Lasso. 
\end{compactitem}

\subsection{Solving TV Regression}
\label{subsec:solve}
The additive model in \eqref{eqn:tvregreg} can be viewed as a linear transform on the model parameters $u_0$ and $u := [u_1; u_2; \dots; u_K]$:
\begin{equation}\label{eqn:additive2}
	\tau = u_0 \vct{1} + A u,
\end{equation}
in which $\vct{1} \in \R^{N_x}$ is all one vector and $A \in \R^{N_x\times |u|}$ is the corresponding linear operator. Here, $N_x = \Pi_{d=1}^D |V_d|$ is the number of all possible covariate combinations for $x$. TV in \eqref{eqn:tvregreg} can be rewritten as
\begin{equation}\label{eqn:TV2}
	TV(\tau) = \|D u\|_1 = \sum_{k=1}^K w_k \|D_k u_k\|_1,
\end{equation}
where $D_k$ is the (sparse) matrix which encodes the edges and $\ell_1$ penalty in~\eqref{eqn:tvfor1st} and \eqref{eqn:tvforhighorder}, and $D$ encodes all the information including the weights. With $M := \text{diag}\{M_e(x)\}$, the TV regularized regression~\eqref{eqn:tvregreg} can be rewritten as
\begin{equation}\label{eqn:tvregreg2}
\min_{u_0, u} \quad  \frac{1}{2} (u_0 \vct{1} + A u - \hat{\tau})^T M (u_0 \vct{1} + A u - \hat{\tau}) +  \lambda \|D u\|_1.\\
\end{equation}
The main problem is to solve $u$ and then the constant term $u_0$ follows trivially. Define $\mtx{P}_{\vct{1}} = \frac{\vct{1} \vct{1}^T M}{\vct{1}^T M \vct{1}}$, $B = A^T M (I - \mtx{P}_{\vct{1}})A$, $b = A^T M (I - \mtx{P}_{\vct{1}}) \hat{\tau}$ and $c = \hat{\tau} M (I - \mtx{P}_{\vct{1}}) \hat{\tau}$. \eqref{eqn:tvregreg2} can be decomposed as first solve
\begin{equation}\label{eqn:tvregreg3}
\min_{u} \quad  \frac{1}{2}(u^T B u - 2 b^T u + c) + \lambda \|D u\|_1,
\end{equation}
and then $u_0 = \frac{\vct{1}^T M (\hat{\tau} - A u)}{\vct{1}^T M \vct{1}}$.

We solve \eqref{eqn:tvregreg3} using an ADMM algorithm. Details are put in the Appendix. 

\subsection{How to Choose Weights?}
\label{subsec:weights}
It is crucial to choose the correct weights $\{w_k\}_{k=1}^K$ in our total variation~\eqref{eqn:tvforadditive}. We provide a data-adaptive method which chooses weights based on the training data alone without a separate test set and expensive parameter tuning. This method of choosing weights can be also applied to other problems, such as group lasso \citep{yuan2006model} with/without overlap groups. 

The basic idea is that the weights $\{w_k\}_{k=1}^K$ for corresponding additive terms $u_k$ should play the role of balancing the penalties such that they don't over-penalize any term because the number of edges is larger. To achieve that, we propose a thought experiment. If there is no HTE, i.e. $u = 0$ in \eqref{eqn:tvregreg2}, then as we decrease $\lambda$ from $\infty$ to 0, there exists a range of $\lambda$ that the solution of \eqref{eqn:tvregreg2} contains nonzero $u_k$. We know these are noises and the whole purpose of penalty terms is to suppress them. In the same line of thinking, if our choice of $\{w_k\}_{k=1}^K$ is ``fair'', then those noises should show up at the same $\lambda$ in the path. If $u_d$ keeps turning into nonzero before $u_f$, then we over-penalized $u_f$ compared to $u_d$. In other word, we choose weights such that when we decrease $\lambda$ all $u_d$ will join the solution path at the same time (on average). 

\begin{theorem}\label{thm:tvregzero} 
 The unique minimizer of \eqref{eqn:tvregreg3} is $u = 0$ if and only if 
    \begin{equation}\label{eqn:tvregzero}
        \sup_{u \neq 0} \frac{u^T b}{\|D u\|_1} \le \lambda
    \end{equation}
Let $b = [b_1; b_2; \dots; b_K]$ and with $\|D u\|_1 = \sum_{k} w_k \|D_k u_k\|_1$, $u = 0$ if and only if
    \begin{equation}\label{eqn:tvregzero2}
        \max_{1\le k \le K}\sup_{u_k \neq 0} \frac{u_k^T b_k}{w_k \|D_k u_k\|_1} \le \lambda.
    \end{equation}
Let $k^*$ be the $k$ that $\sup_{u_k \neq 0} \frac{u_k^T b_k}{w_k \|D_k u_k\|_1} = \lambda$, $u_{k^*}$ joins the solution path at $\lambda$.
\end{theorem}
Theorem~\ref{thm:tvregzero} can be proved using Lasso dual, as in \citep{hastie2015statistical}. Let
\begin{equation}\label{eqn:dual}
    \gamma_k = \sup_{u_k \neq 0} \frac{u_k^T b_k}{\|D_k u_k\|_1},
\end{equation}
and left hand side of \eqref{eqn:tvregzero2} can be simplified as $\max_{1\le k \le K} \frac{\gamma_k}{w_k}$. Recall $b = A^T M (I - \mtx{P}_{\vct{1}}) \hat{\tau}$ is a random vector. We pick $\{w_k\}_{k=1}^K$ such that $\gamma_k/w_k$ are all the same in expectation when $\hat{\tau}$ is pure random noise:
\begin{equation}\label{eqn:weights}
	w_k \propto \Expect(\gamma_k) \quad \forall k = 1,2,\dots, K,
\end{equation}
where $b_k$ is the k\ts{th} entry of the random vector $A^T M (I - \mtx{P}_{\vct{1}}) \vct{n}$ where $\vct{n}\sim \mathcal{N}(0, M^{-1})$.

\subsection{Parameter Tuning for $\lambda$}
\label{subsec:lambdatuning}
When $\lambda$ is sufficiently large, the solution of our TV regression is 0, as shown in Theorem~\ref{thm:tvregzero}. As $\lambda$ decreases, more and more heterogeneous treatment effects come out, and finally $u$ becomes the standard least square solution when $\lambda = 0$. In other words, the model overfits the data when $\lambda$ is too small, but suppress all signals when $\lambda$ is sufficiently large, see Theorem~\ref{thm:tvregzero}. It is important to stop at the right $\lambda$: the one that ideally captures the true heterogeneous treatment effects and suppress the false positives. 

We achieve these two goals by simply solving Eqn.~\eqref{eqn:tvregreg3} with multiple $\lambda$'s in a descending order. This can be very efficient because we can use the solution for the last $\lambda$ as the initial parameter for the next $\lambda$ and the two solutions should be very close, see \citep{hastie2015statistical}. For $\lambda$ selection, to avoid computational expansive alternatives such as cross validation or a separate test set. We use information criteria such as AIC \citep{akaike1998information} and BIC \citep{schwarz1978estimating}. Since TV Regression, like Lasso, will introduce bias, for each $\lambda$, after clusters have been discovered by TV Regression, we fit an OLS (ordinary least square) using only useful additive terms without the TV penalty in~\eqref{eqn:tvregreg3}, and record its residual $Res(\lambda)$ (weighted by $M_e(x)$ at each $x$). This model re-fitting is similar to Lasso de-biasing and is a common practice. We can define AIC and BIC of the de-biased model as 
\begin{align}
    BIC(\lambda) &= 2 Res(\lambda) + Dof(\lambda) \log(N_t).\label{eqn:bic}\\
    AIC(\lambda) &= 2 Res(\lambda) + 2 Dof(\lambda)\label{eqn:aic}
\end{align}
where $Dof(\lambda)$ is the degree of freedom of the linear model and equals 1 plus the number of additive terms in the linear model, $N_t$ is the number of points where $M_e(x) > 0$ (pixels in the ``image''). We pick the $\lambda$ (or the associated model) which has the smallest information score. As a side benefit from fitting the OLS, we can also perform significant test \citep{allofstat} on each effect terms to decide whether they might be false signals.

\subsection{Post TV Regression Reprocess}\label{subsec:posttv}
TV regression help us to identify block-wise clusters that can be used to concisely summarize the HTE $\tau(x)$. These identifies clusters can then be treated as building blocks and used as new features to reprocess the data so that might be able to further improve conciseness of the representation. To see that, suppose there are two clusters A and B for browser and there exist a global effect 0 and 0.1 on cluster A and 0 on B. When A and B have the same size, we can see a solution of 0.05 global effect, 0.05 for A and -0.05 for B has the same TV penalty as the ground truth. If we define concise as the smallest number of HTE effects we need to use to summarize the structure, then TV regression might fail to identify the most concise solution. Similarly when a ground truth second order effect covers a big portion of the population, TV regression might identify it as 2 first order effects and 1 second order effect (every second order effect can equivalently be represented by one first order effect subtracts another first order effect and adds one second order effect.). A post reprocess also means TV regression solution is allowed to contain some false positive effect terms as long as all true block structures are captured, giving us more flexibility and robustness so that $\lambda$ tuning does not need to be perfect. 

Minimizing the number of effects used to describe HTE can be achieved by a linear model with $\ell_0$ penalty. We take all first and second order clusters and its complements identified in TV Regression. Note that complement of a second order block is not itself a block, we need to further break it into blocks. We fit \eqref{eqn:master} using these first order cluster and second order blocks as predictors with an ElasticNet penalty \citep{zou2005regularization,mazumder2012sparsenet} that closely approximates $\ell_0$ penalty. An additional step of fitting an OLS using only nonzero terms in the result of ElasticNet can be used to debias the estimated effects and also provide confidence intervals and significant tests. 

Confidence intervals and statistical tests in the above OLS doesn't take into account the fact that the model is from careful model selection using TV Regression and ElasticNet. A more recent area of research aims to provide correct post-selection inference \citep{taylor2014post,taylor2015statistical} and also a more principled way to search best model in a Lasso-like solution path. Results there can also be applied to TV Regression since it belongs to the family of generalized Lasso. 

\subsection{Multiplicative Effect}
\label{subsec:multiplicative}
It is very common in practice that the relative effect (percent change) ${y^{(1)}}/{y^{(0)}}-1$ is of interest, we can fit an multiplicative model for ${y^{(1)}}/{y^{(0)}}$, i.e.,
\begin{equation}\label{eqn:multiplicative}
	\frac{y^{(1)}}{y^{(0)}} = v_0 \Pi_{k=1}^K v_k(x).
\end{equation}
Taking log on both sides, we can apply our TV regression~\eqref{eqn:tvregreg} to $\log(y^{(1)}/y^{(0)})$, i.e.,
\begin{equation}\label{eqn:multiplicative2}
	\log\left(\frac{y^{(1)}}{y^{(0)}}\right) = u_0 + \sum_{k=1}^K u_k(x).
\end{equation}
Now the measurement at $x$ are $\log(\bar{y}^{(1)}(x)/\bar{y}^{(0)}(x))$ whose variance can be estimated by the Delta Method \citep{asympstat} as:
\begin{equation}\label{eqn:variance2}
	\hat{\Var}_{\log}(x) = \frac{\Var^{(0)}(x)}{(\bar{y}^{(0)})^2} + \frac{\Var^{(1)}(x)}{(\bar{y}^{(1)})^2}.
\end{equation}
The matrix $M = \text{diag}\{1/\hat{\Var}_{\log}(x)\}$ can be computed correspondingly.

\subsection{Signal Importance Score}\label{subsec:score}
We define importance of an additive term by comparing the prediction error with and without an extra term. Typically prediction error need to be evaluated on a hidden test set. We use Stein's unbiased estimator which does not require testing on a separate test set. 

\subsection{Covariate Pre-Screening}\label{subsec:prescreen}
When the number of covariate is large, the number of covariate combinations can be larger than the original sample size N. Moreover, the optimization problem \eqref{eqn:master} involves $M_e(x)$ which relies on reasonable variance estimations of $var^{(w)}(x), w=0,1$. When sample sizes $N^{(w)}(x),w=0,1$ are small, variance estimation and effect estimation $\hat{\tau}(x)$ can be unstable. To overcome that, we can simply use a Group Lasso \citep{yuan2006model} step to first screen out irrelevant covariates. This pre-screening stage does not need to achieve effect clustering, which is the task for the following TV regression on a much smaller set of covariates.  



\section{Empirical Results}
\label{sec:examples}
\begin{figure}[!htb]
    \centering
    \includegraphics[width=0.45\textwidth]{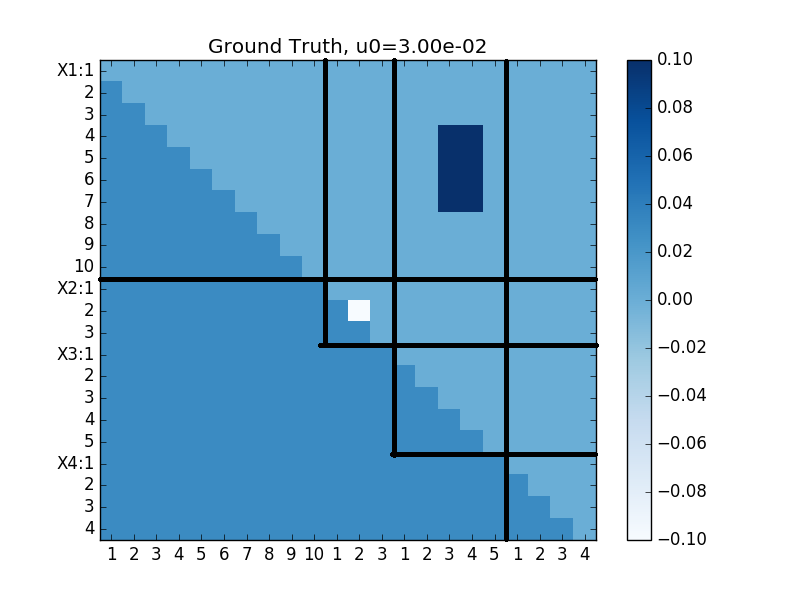}
    \caption{Example~\ref{ex:1}: ground truth effects in a matrix heatmap. Covariate names for y-axis is omitted. }
    \label{fig:synthetic_truth}
\end{figure}

In this section we apply TV Regression to both synthetic data and also real experiment data. Throughout this section we visualize a solution of TV Regression using a 2d matrix heatmap. This visualization is only for technical illustration and is not the final concise summary. Ground truth solution for Example~\ref{ex:1} is shown in Figure~\ref{fig:synthetic_truth}. 
Recall that the ground truth is global effect 0.03, first order -0.1 on $X_2$:$2$ and second order effect 0.1 on the block  $X_1$$:$$(4,5,6,7)$ $\cap$ $X_3$$:$$(3,4)$. We concatenate all levels of 4 covariates together as both x-axis and y-axis and form a matrix where
\begin{compactenum}
   \item Diagonal entries represents first order effect $u_d(x_d)$,
   \item Upper triangular entries represents second order effect $u_{d,f}(x_d, x_f)$, 
   \item The whole lower triangular entries are the same, representing global effect $u_0$. 
\end{compactenum}
In Figure~\ref{fig:synthetic_truth} the lower triangular displays global effect 0.03. The white pixel on the diagonal line is the -0.1 effect for $X_2$:$2$ and the dark blue block in the upper triangular stands for the 0.1 second order effect. 

\subsection{A synthetic example}
\label{subsec:synthetic}
We run TV-regression to fit a second-order model~\eqref{eqn:secondordermodel} for Example~\ref{ex:1} in Section~\ref{subsec:issue}. For TV penalty, we simply use a complete graph on all 4 covariates assuming they are all categorical without a natural topology. We take $\alpha = 0.5$ in Eqn.~\eqref{eqn:tvfor1st} and \eqref{eqn:tvforhighorder} to equally weight the total variation and $\ell_1$ penalty. We use 10,000 random samples to estimate the weights $w_k$ in Eqn.~\eqref{eqn:weights}. We visualize the solution path (after coefficient debias using OLS for each $\lambda$), as described in Section~\ref{subsec:lambdatuning}. 
\begin{figure}[!htb]
    \centering
    \includegraphics[width=0.225\textwidth]{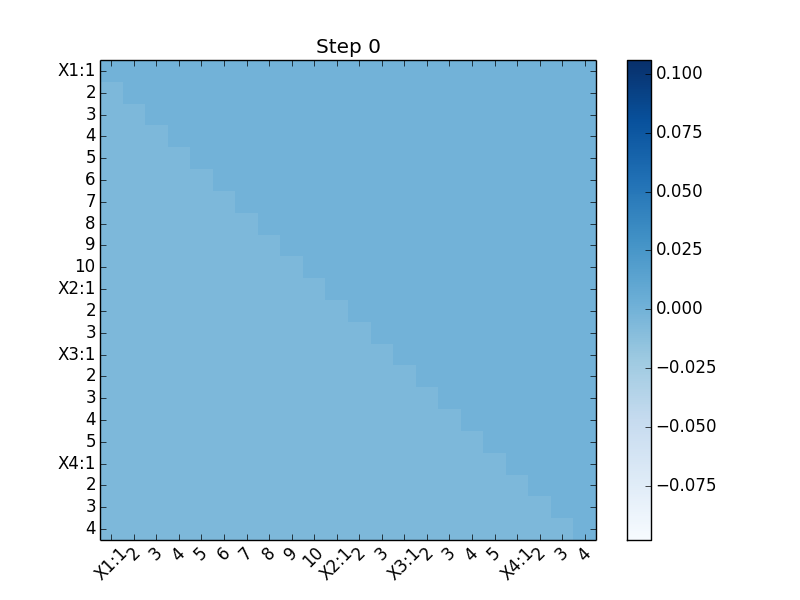}
    \includegraphics[width=0.225\textwidth]{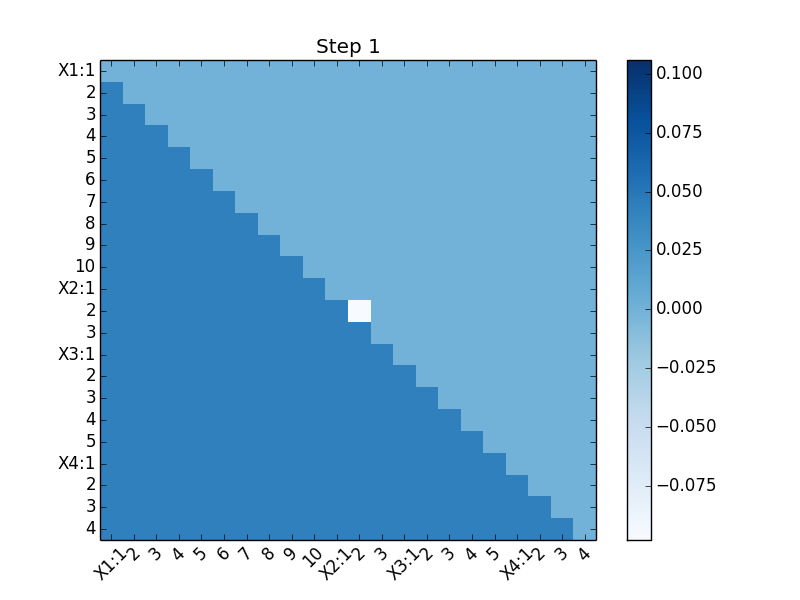}\\
    \includegraphics[width=0.225\textwidth]{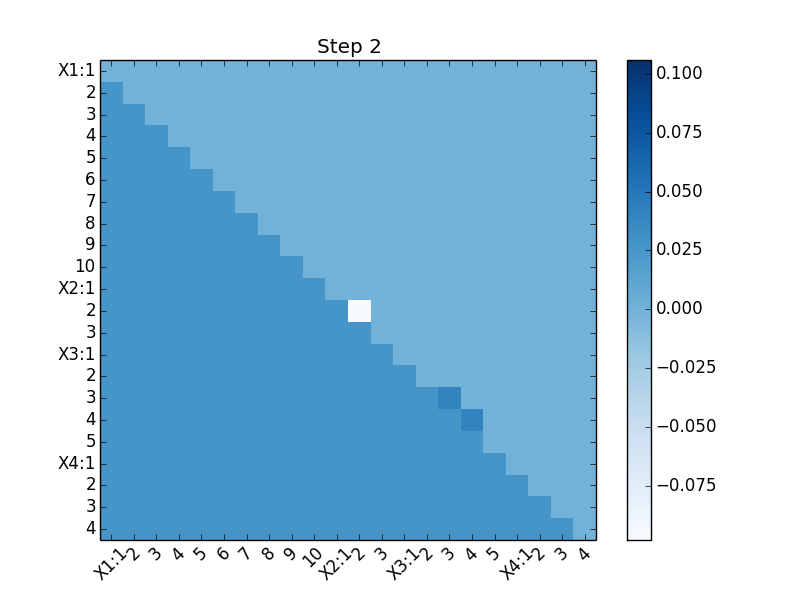}
    \includegraphics[width=0.225\textwidth]{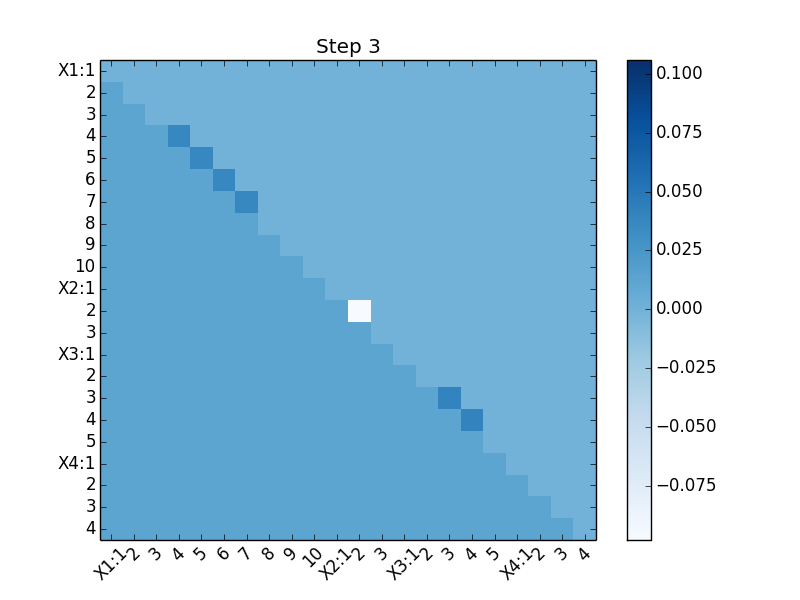}\\
    \includegraphics[width=0.225\textwidth]{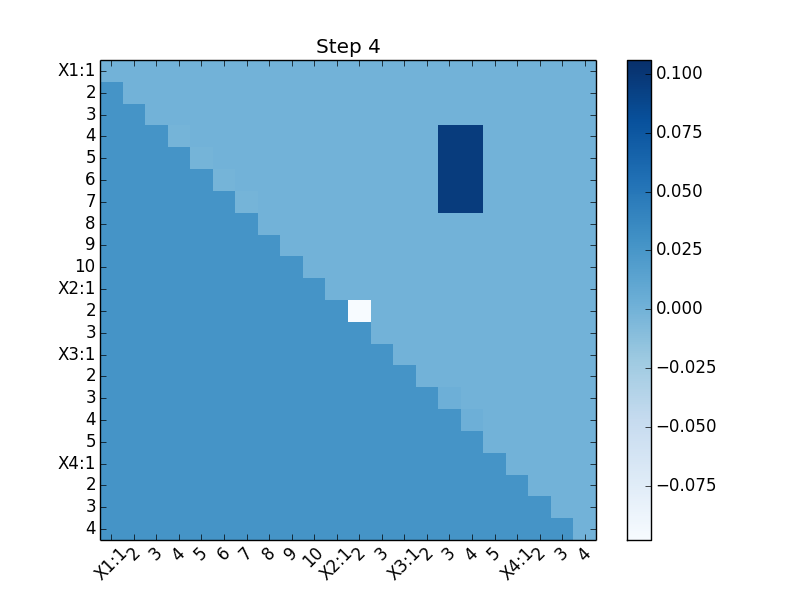}
    \includegraphics[width=0.225\textwidth]{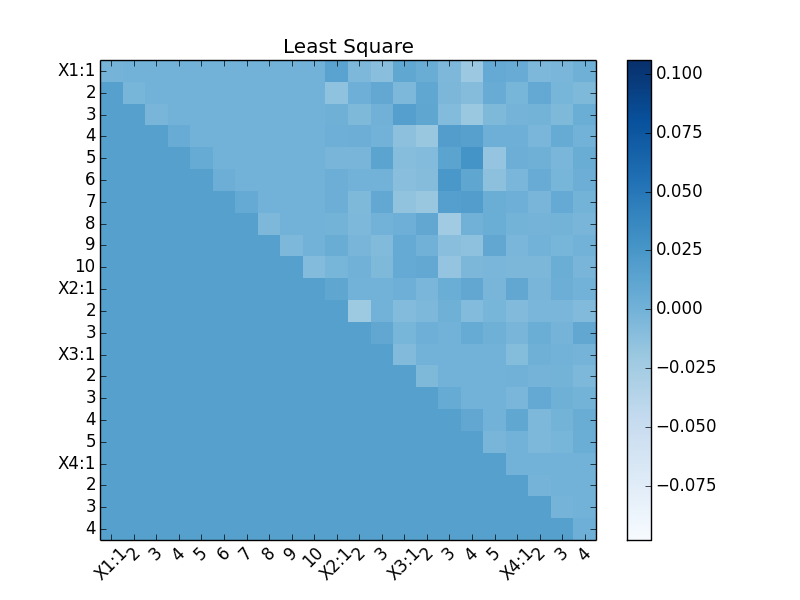}
    \caption{Solution path ordered row-wise from left. The six plots corresponds to $1/\lambda$ values at $0.5$, $0.8$, $2.49$, $4.372$, $4.442$ and $+\infty$. See Table~\ref{tab:synthetic_bic}.}
    \label{fig:synthetic_path}
\end{figure}
The solution path start with only global effect on the top left, first detected first order effect on $X_2:2$ when $\lambda$ decreases. It then detect first order effect $X_3$$:$$(3,4)$ and $X_1$$:$$(4,5,6,7)$ (middle plots). It then figured out a better representation is a second order effect  $X_3$$:$$(3,4)\cap X_1$$:$$(4,5,6,7)$. We point out that the first order effects on X1 and X3 are still nonzero at this stage, but are too small to be visible in Figure~\ref{fig:synthetic_path} bottom left. As we further decrease $\lambda$ the solution converges to the least square solution. 

We tabulate the BIC score for all the models above in Table~\ref{tab:synthetic_bic}. We select the model with $1/\lambda = 4.442$ (Figure~\ref{fig:synthetic_path} bottom left) since it has the smallest BIC score. We can see that the selected model exactly captures the two effects in the ground truth, even though it also contains two very small first order effect on $X_3$$:$$(3,4)$ and $X_1$$:$$(4,5,6,7)$. To further improve the result, we applied a post TV reprocess as described in Section~\ref{subsec:posttv}. The effect of this reprocess is shown in Table~\ref{tab:synthetic_reprecess}. Reprocess removed the two false small first order effects mentioned above, and report an equivalent alternative presentation for the ground truth: instead of reporting the first order effect of -0.1 on $X2$$:$$2$, it reports first order effect of 0.1 on $X2$$:$$(1,3)$, and reduce the global effect by 0.1 to -0.07. They are equally concise by the number of HTE terms. 
\begin{table}[!htb]
\centering 
\begin{tabular}{@{}llllllll@{}}
\toprule
$1/\lambda$ & 0.5  & 0.8  & 2.49 & 4.372 & 4.442 & 7   & 14  \\ \midrule
\# Effects  & 0    & 1    & 2    & 3     & 4     & 5   & 7   \\ \midrule
BIC         & 4283 & 2574 & 1492 & 1159  & 632   & 638 & 644 \\ \bottomrule
\end{tabular}
\caption{The solution path of Example~\ref{ex:1}} 
\label{tab:synthetic_bic} 
\end{table}

\begin{table*}[!tb]
\centering
\begin{tabular}{@{}lllllllll@{}}
\toprule
              & global & X1:(4,5,6,7) & X1:(1,2,3,8,9,10) & X2:2   & X2:(1,3) & X3:(3,4) & X3:(1,2,5) & X1:(4,5,6,7)$\cap$X3:(3,4) \\ \midrule
TV Regression+OLS & 0.027  & -0.0009*     &                   & -0.098 &          & 0.0026*  &            & 0.096                      \\
Elastic Net+OLS   & -0.072 &              &                   &        & 0.101    &          &            & 0.103                      \\ \bottomrule
\end{tabular}
\caption{Two weak first order effects(*) in TV Regression result are removed after the reprocess. They also do not pass t-test in the OLS fit (p-value 0.66 and 0.15). All other signals pass t-test in OLS with very small p-values. Reprocessing with Elastic Net found an equivalent representation with the same number of HTE term as the ground truth. }
\label{tab:synthetic_reprecess}
\end{table*}

\subsection{TV-regression vs. Lasso}
\label{subsec:comparelasso}
\begin{figure}[!htb]
    \centering
    \includegraphics[width=0.225\textwidth]{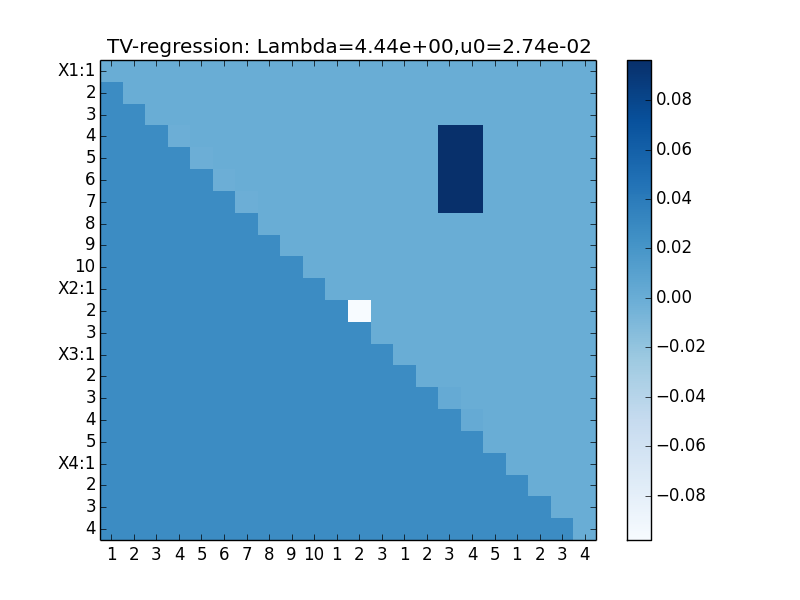}
    \includegraphics[width=0.225\textwidth]{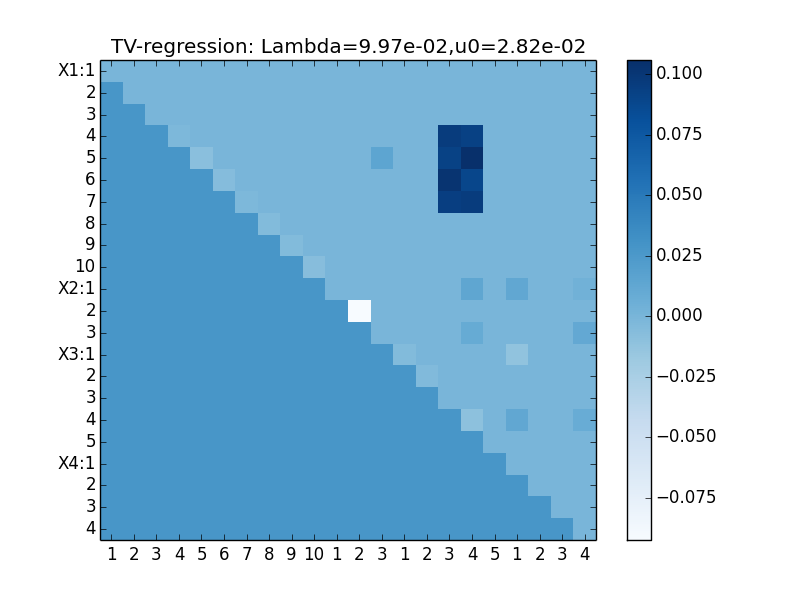}
    \caption{Comparison between TV-regression with $\alpha = 0.5$ (left) and Lasso (right, TV-regression with $\alpha = 1$). Both models are selected by BIC.}
    \label{fig:synthetic_comp}
\end{figure}

\begin{figure}[!htb]
    \centering
    \includegraphics[width=0.225\textwidth]{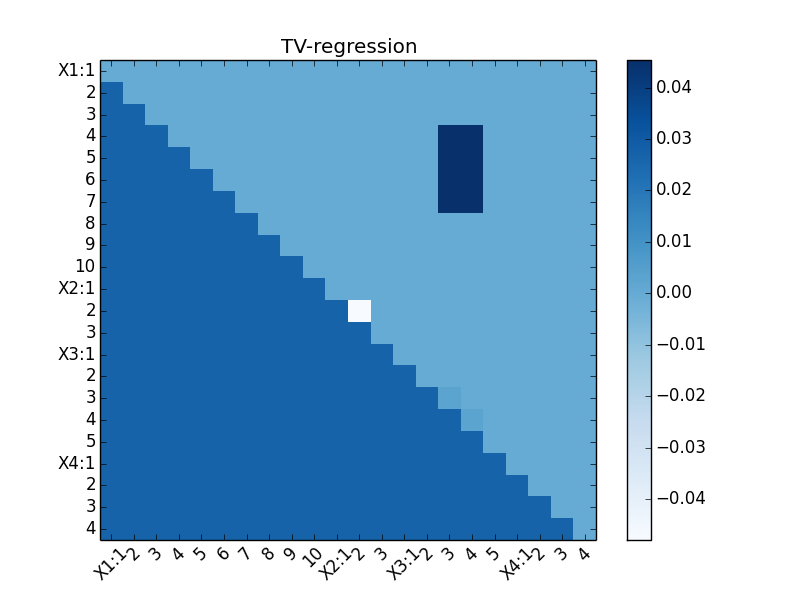}
    \includegraphics[width=0.225\textwidth]{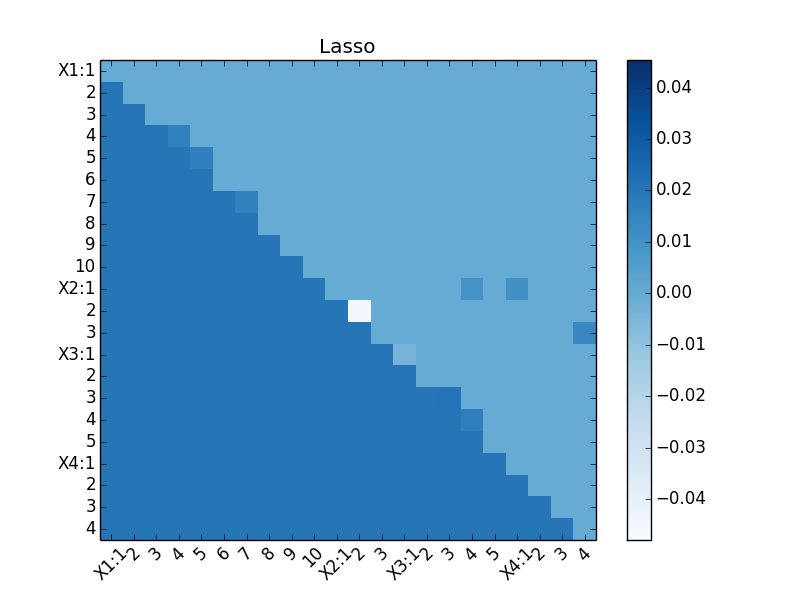}
    \caption{Comparison between TV-regression with $\alpha = 0.5$ (left) and Lasso (right, TV-regression with $\alpha = 1$) with ground truth effects reduce to $\pm0.05$ from $\pm0.1$.}
    \label{fig:synthetic_comp_weak}
\end{figure}

We also obtained the lasso path ($\alpha = 1$ in \eqref{eqn:tvforhighorder}) for Example~\ref{ex:1} and compared it to that of TV-regression with $\alpha = 0.5$. The BIC selected models are shown in Figure~\ref{fig:synthetic_comp}.
We saw that
\begin{compactenum}
\item Although Lasso captures both first and second order effects, it wasn't designed to discover the block structure of the second order effect. 
\item Lasso also contains more false signals than TV Regression. We believe this is still true if cross-validation is used for $\lambda$ tuning, as reported in \citep{tian2014simple}.
\end{compactenum}
The true signal in Example~\ref{ex:1} is actually relatively strong comparing to the noise. If we lower the strength of signal from 0.1 and $\pm0.1$ to $\pm0.05$ and run the same example, as shown in Figure~\ref{fig:synthetic_comp_weak}, we found Lasso completely missed the second-order block with many false signals! In comparison, TV regression still captures the group truth, with only one very weak false signal on $X_3$$:$$(3,4)$ which is later removed in post reprocess. 

\subsection{An experiment on a personalized recommendation}
\label{subsec:rec}
\begin{figure}[!htb]
    \centering
    \includegraphics[width=0.48\textwidth]{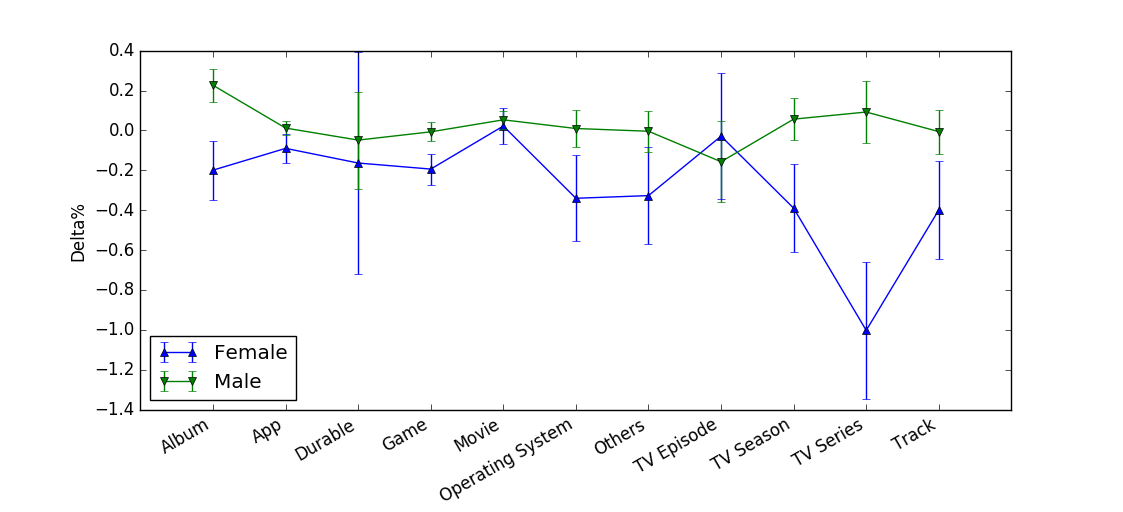}
    \caption{Example~\ref{ex:2}: mean and standard deviation of the percent change by gender and product type.}
    \label{fig:rec_data}
\end{figure}

\begin{ex}\label{ex:2}
Microsoft ran an online controlled experiment to evaluate a new gender based personalized recommendation algorithm. We have two covariates (Gender and ProductType) in this example. The mean and standard deviation ($1/\sqrt{M_e}$ in Eqn.~\eqref{def:M}) of the percent change for an important metric are shown in Figure~\ref{fig:rec_data}.
\end{ex}

\begin{figure}[!htb]
    \centering
    \includegraphics[width=0.225\textwidth]{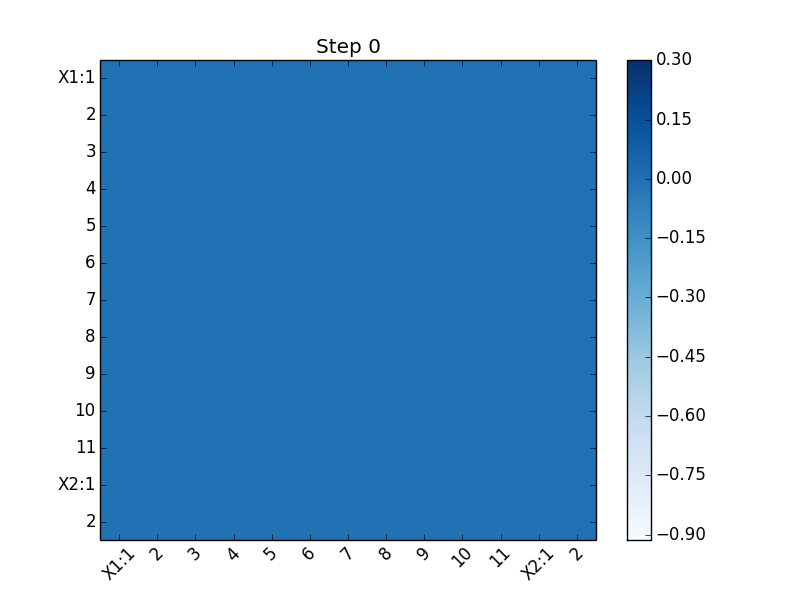}
    \includegraphics[width=0.225\textwidth]{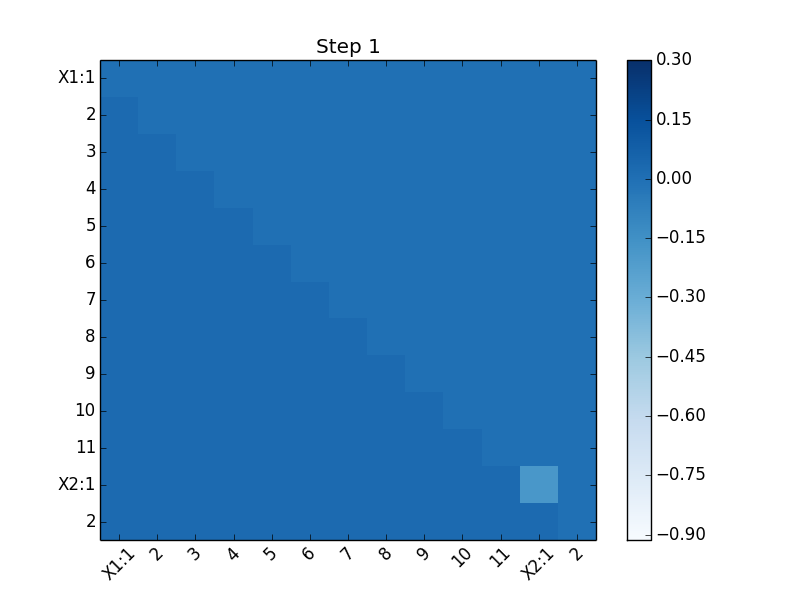}\\
    \includegraphics[width=0.225\textwidth]{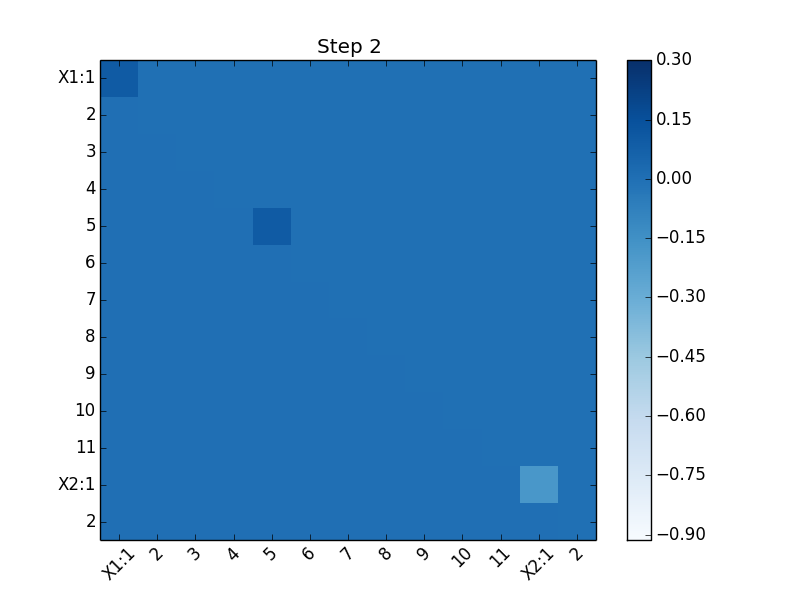}
    \includegraphics[width=0.225\textwidth]{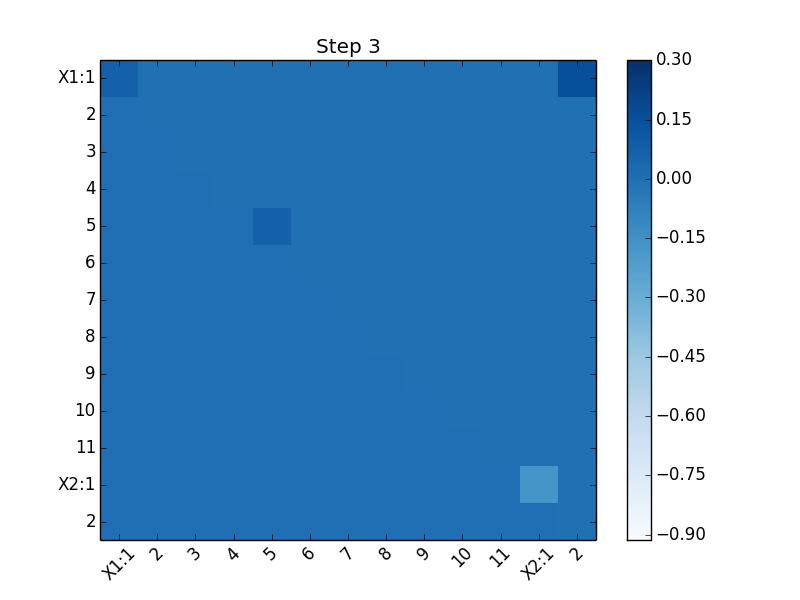}\\
    \includegraphics[width=0.225\textwidth]{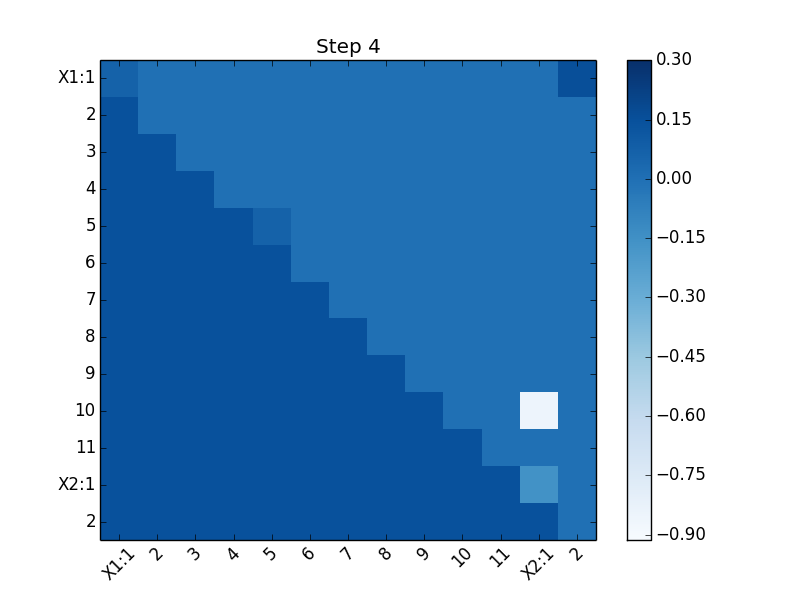}
    \includegraphics[width=0.225\textwidth]{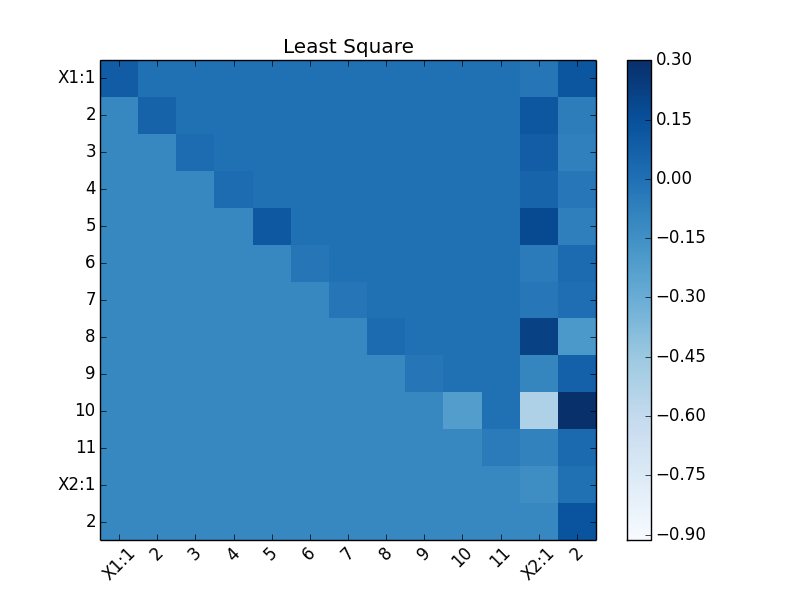}
    \caption{Solution path of TV Regression for Example~\ref{ex:2}. Bottom left: $1/\lambda =  5.8$ (smallest BIC); bottom right: $1/\lambda = \infty$ (least square).}
    \label{fig:rec_path}
\end{figure}

We again use complete graph for both covariates since they are unordered. We run TV-regression with $\alpha=0.5$, and the first five models in the solution path and the least square solution ($1/\lambda=\infty$) are shown in Figure~\eqref{fig:rec_path}.

We select the model with $1/\lambda = 5.8$ (Figure~\ref{fig:rec_path} bottom left) using BIC. The selected model contains two first order effects: strong Gender:Female effect and a weaker ProductType:Album\&Movie effect together with two second order effects: (Male, Album) and (Female, TV series). All the selected effects aligns well with the Figure~\ref{fig:rec_data} where there is certainly a first order effect for gender and female customers have more negative effect for Album and TV Series. The weaker first order ProductType:Album\&Movie effect has a p-value of 0.04 in the post reprocess stage and therefore is a boarder-line effect. 

\subsection{Diagnosis in KPI Monitoring}\label{subsec:bing}
\begin{ex}\label{ex:3}
We observed the revenue per volume (RPV) of Product A for the recent week is lower than the week(s) before. We suspect this is not a systematic decline but rather a potential problem of a subpopulation. For illustration, we choose four covariates (DeviceOSName with 5 levels, DeviceType with 2 levels, ProductType with 13 levels and WeekDays with 7 levels). Real application was run on a large number of covariates so a pre-screening stage described in Section~\ref{subsec:prescreen} was used. 
\end{ex}
The gross RPV can be calcuated from the RPV of subpopulations as below:
\begin{equation}\label{eqn:rpv}
    \text{RPV} = \frac{\sum_{i} \text{rpv}_{i} \text{vol}_i}{\sum_i \text{vol}_i} = \sum_{i} \text{rpv}_{i} \text{w}_i,
\end{equation}
where $w_k = \text{vol}_{i}/\sum_i \text{vol}_i$ is the weight of volumes from subpopulation $i$. 

\begin{figure}[!htb]
    \centering
    \includegraphics[width=0.225\textwidth]{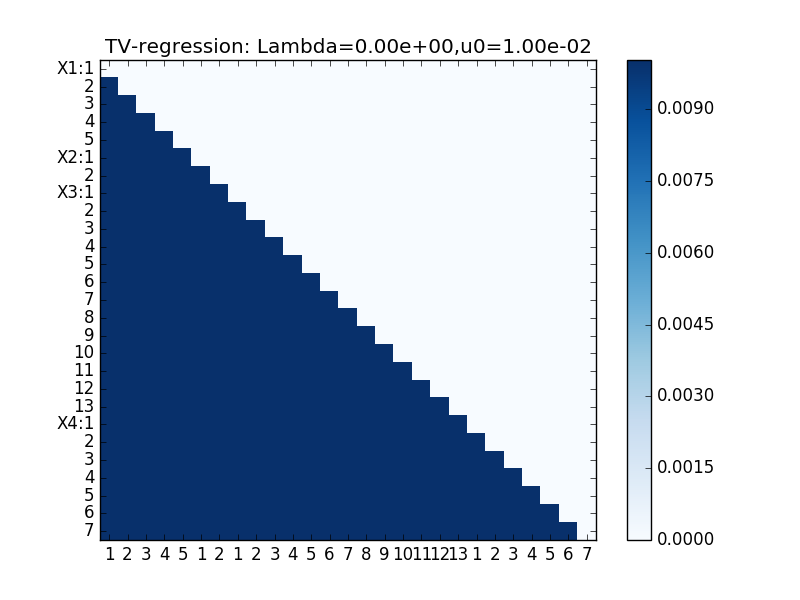}
    \includegraphics[width=0.225\textwidth]{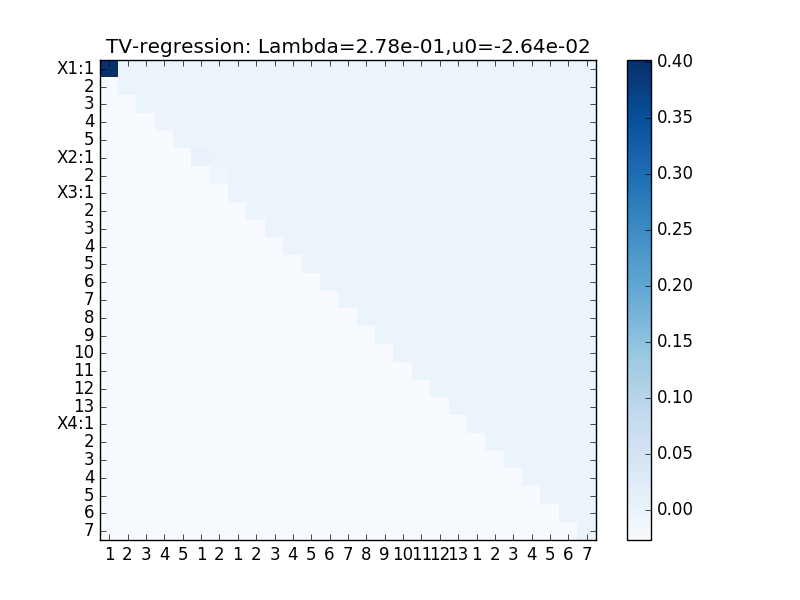}
    \caption{TV Regression results for percent changes of revenue per volume $\text{rpv}_i$ (left) and weights $w_i$ (right).}
    \label{fig:ex3}
\end{figure}

For covariates DeviceOSName, DeviceType and ProductType, we simply use complete graph to define the total variation. For WeekDays, we use the loopy graph in which two consecutive weekdays (Monday and Tuesday, Tuesday and Wednesday, ..., and Sunday to Monday) are connected. We compute the mean and standard deviation of the percent change treating pre-period as control and post-period as treatment for both $\text{rpv}_{i}$ and $w_i$, as in Example~\ref{ex:2}. We then run TV-regression with $\alpha=0.5$. Figure~\ref{fig:ex3} shows the BIC selected models.

The results in Figure~\ref{fig:ex3} (left plot) suggested no HTE at all for revenue per volumne. For weights $w_i$ (Figure~\ref{fig:ex3} right), it showed a strong increase (40\%) on the first level of DeviceOSName. In fact, the first level of DeviceOSName was Android and it turned out that the volume from one device with Android system increased a lot in the post-period. Also, comparing to devices with other operation systems, Android devices had a relatively low rpv. Therefore, the volume increase in this low-rpv Android device caused the drop of the gross RPV.

\subsection{Adaptive Between-Graph TV Weights}
\label{subsec:weightexample}
\begin{figure}[!htb]
    \centering
    \includegraphics[width=0.225\textwidth]{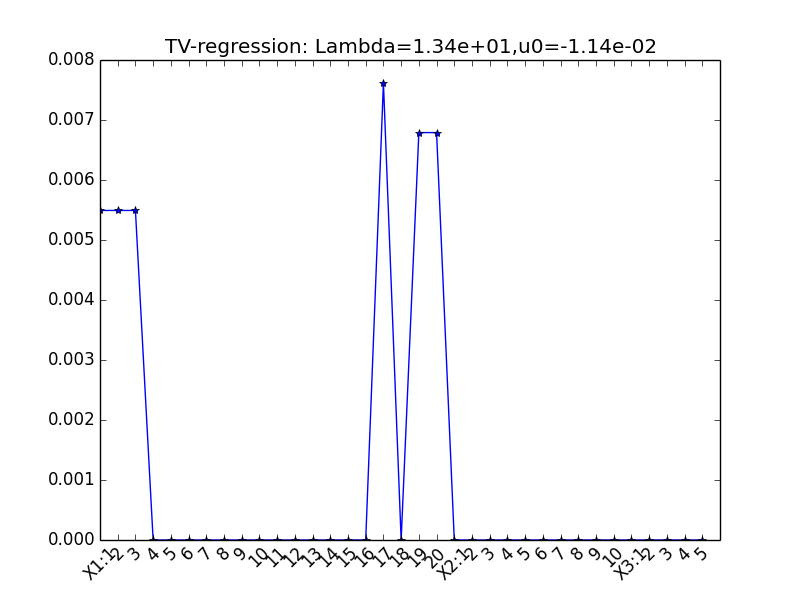}
    \includegraphics[width=0.225\textwidth]{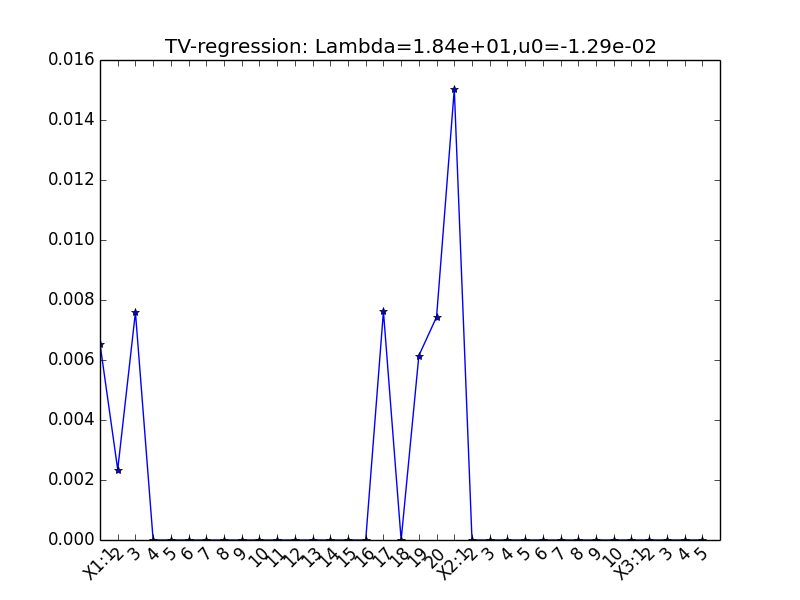}
    \caption{Part of the solution path with equal weights. Left: $1/\lambda =  13.4$; Right: $1/\lambda =  18.4$. BIC chose the model on the right.}
    \label{fig:ex4_noweights}
\end{figure}

We demonstrate the necessity of adaptive between-graph weights $w_k$ in the definition of total variation~\eqref{eqn:tvforadditive} using a synthetic example. We described our procedure in Section~\ref{subsec:weights}.
\begin{ex}\label{ex:4}
There are 3 categorical covariates $X_1,X_2, X_3$. $X_1$ is an ordered covariate with 20 equally weighted levels. Its graph connects every consecutive pair. $X_2$ has 10 equally weighted levels. $X_3$ has 5 levels, with proportion 10\%, 30\%, 30\%, 20\% and 10\%. Both $X_2$ and $X_3$ have a complete graph. We put a simple ground truth effect as this: $X_2$$:$$1$ has a first order effect of 0.015 and there's no other heterogeneous effects. In addition, there is a constant background effect -0.01. $10,000$ \textit{i.i.d.} treatment and control observations are simulated from $N(0, 0.1)$ respectively with treatment effect added to treatment observations according to the prescribed design.
\end{ex}
TV-regression with $\alpha = 0.5$ and a first order additive model is applied to this example. First we simply use equal weights, i.e. $w_1 = w_2 = w_3 = 1$ in~\eqref{eqn:tvforadditive}, part of the obtained solution path is shown in Figure~\ref{fig:ex4_noweights}. The BIC picks the model on the right in Figure~\ref{fig:ex4_noweights}. Note that $X_2$ and $X_3$ are categorical so in Figure~\ref{fig:ex4_noweights} only the lines connecting $X_1$'s levels mean ordering. 

\begin{figure}[!htb]
    \centering
    \includegraphics[width=0.225\textwidth]{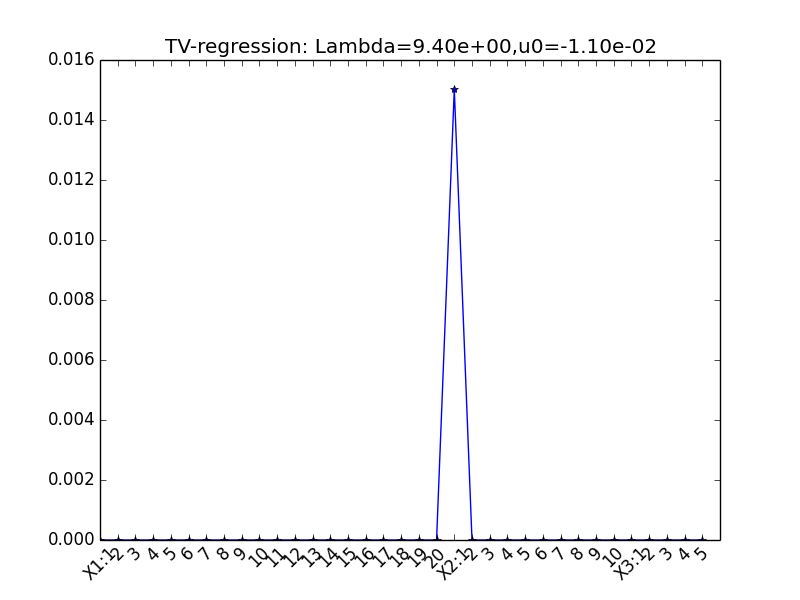}
    \includegraphics[width=0.225\textwidth]{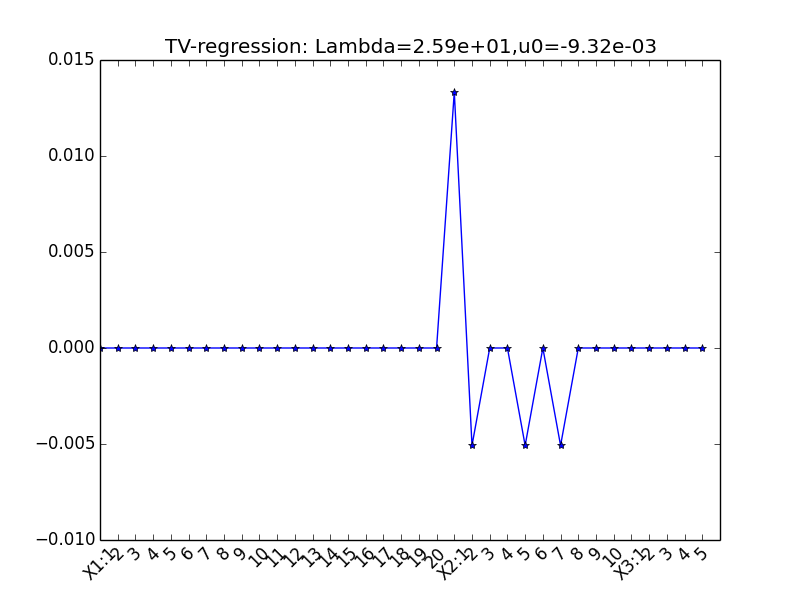}
    \caption{Solution path with weights~\eqref{eqn:weights}. Left: $1/\lambda = 9.40$; right: $1/\lambda = 25.9$.}
    \label{fig:ex4_weights}
\end{figure}

Although the selected model contains the ground truth effect on $X_2$:$1$, there are many false positive effects due to noise. The reason is because $X_1$ is ordered, and there is considerably less edges in its TV graph compared to $X_2$ and $X_3$. If we use equal weights between the graphs, then false positive in $X_1$ will be less penalized than those in $X_2$ and $X_3$. 
In comparison, we use Monte Carlo method with 10,000 samples to estimate the weights in Eqn.~\eqref{eqn:weights}. The results are
\begin{equation*}
    (w_1, w_2, w_3) = (88.5, 15.4, 32.5).
\end{equation*}
As expected, we put higher weights on $X_1$ and less on $X_2$ and $X_3$. $X_3$ got higher weights than $X_2$ because is only has 5 levels compared to $X_2$ having 10. The corresponding solution path is shown in Figure~\ref{fig:ex4_weights}. The BIC picks the model on the right side in Figure~\ref{fig:ex4_weights}.
We can see that 
\begin{compactenum}
\item The selected model contains the true positive effect on $X_2$:$1$ and one false positive effect on $X_2$:$(2,5,7)$.
\item The ground truth effect on $X_2$:$1$ enters the solution path in the first place and has higher importance score than $X_2$:$(2,5,7)$.
\end{compactenum} 
Therefore, TV-regression with weights~\eqref{eqn:weights} serves our goals better than TV-regression with equal weights. We also ran post reprocess. In this example, we weren't able to remove the false positive, which has a p-value of 0.001. 

\section{Conclusion and Future works}\label{sec:con}
In this paper we presented Total-Variation regularized regression to achieve simultaneous heterogeneous regression and concise summarization. We bet on sparsity of effects and use an additive model so that HTE can be summarized using a small number of lower order effects. At the same time Total-Variation penalty will further cluster similar effects into block structures, which can be easily interpreted and are memorable as a set of AND clause. We described how to define Total-Variation penalty for both ordered and unordered discrete covariates in detail and proposed a novel method of tuning between-graph weights. 4 examples using both synthetic and real experiment data were used to demonstrate TV regression and compare it to existed approaches such as Lasso. 

One line of future work is to use more rigorous post selection inference to assess the strength of discovered effects and further remove false positives. We had some success using post-TV reprocess, but the t-test we used ignored the fact that the model is selected and therefore still produces more false positive than prescribed significant level. 

Another line of future work is to compare TV regression to alternative fusion penalties such as OSCAR \citep{bondell2008simultaneous}, GOSCAR as well as other non-convex penalties \citep{yang2012feature}. 
\setlength{\bibsep}{0.0pt}
\bibliographystyle{agsm_nourl}
\bibliography{library}

\appendix
\section{An ADMM algorithm}
\label{subsec:ADMM}
We propose to solve Eqn.~\eqref{eqn:tvregreg3} using the alernating direction method of multipliers (ADMM)~\cite{boyd2011distributed}. Consider the unconstrained optimization problem in Eqn.~\eqref{eqn:tvregreg3}, which is equivalent to the following constrained optimization problem:
\begin{equation}\label{eqn:tvregreg3_constrain}
\begin{split}
    \min_{u, z} \quad &\frac{1}{2}(u^T B u - 2 b^T u + c) + \lambda \|z\|_1 \\
    \text{s.t.} \quad &D u = z,
\end{split}
\end{equation}
where $z$ are slack variables. Eqn.~\eqref{eqn:tvregreg3_constrain} can be solved by ADMM. The augmented Lagrangian is
\begin{equation*}
\begin{split}
    L_{\rho}(u, z, y) = &\frac{1}{2}(u^T B u - 2 b^T u + c) + \lambda \|z\|_1 \\
    &+ y^T (D u - z) + \frac{\rho}{2}\|D u - z\|_2^2,
\end{split}
\end{equation*}
where $y$ is the augmented Lagrangian multipliers. The update rule for ADMM consists of the following three steps.

Update $u$: In the (n+1)-th iteration, $u^{n+1}$ can be updated by minimizing $L_{\rho}$ with $z, y$ fixed:
\begin{equation}\label{eqn:uupdate}
   u^{n+1} = \argmin_{u} \frac{1}{2}(u^T B u - 2 b^T u + c) + (y^n)^T D u + \frac{\rho}{2}\|D u - z^n\|_2^2.
\end{equation}
The above optimization is quadratic, and its solution is given by $u^{n+1} = F^{-1} b^{n}$, where
\begin{equation}\label{eqn:Fandb}
\begin{split}
    F &= B + \rho D^T D, \\
    b^{n} &= b - D^T y^n + \rho D^T z^n.
\end{split}
\end{equation}
The computation of $u^{n+1}$ involves solving a linear system, which is the most time-consuming part in the whole algorithm. To compute $u^{n+1}$ efficiently, we compute the Cholesky decomposition of $F$ at the beginning of the algorithm:
\begin{equation*}
    F = R^T R.
\end{equation*}
Note that $F$ is a positive definite matrix, and is independent of the parameter $\lambda$. Therefore, when computing a solution path, we only need to compute the above Cholesky decomposition once. Using the Cholesky decomposition we only need to solve the following two linear systems at each iteration:
\begin{equation}\label{eqn:uupdate2}
   R^T \hat{u} = b^n, \quad R u^{n+1} = \hat{u}.
\end{equation}

Update $z$: $z^{n+1}$ can be obtained by solving 
\begin{equation*}
   z^{n+1} = \argmin_{z} \frac{\rho}{2}\|D u^{n+1} - z\|_2^2 + \lambda \|z\|_1 - (y^{n})^T z,
\end{equation*}
which is equivalent to the following problem:
\begin{equation}\label{eqn:zupdate}
   z^{n+1} = \argmin_{z} \frac{1}{2}\|z - Du^{n+1} - \frac{1}{\rho} y^{n}\|_2^2 + \frac{\lambda}{\rho} \|z\|_1.
\end{equation}
Eqn.~\eqref{eqn:zupdate} has a closed-form solution, known as soft-thresholding:
\begin{equation}\label{eqn:zupdate2}
   z^{n+1} = S_{\lambda/\rho}(Du^{n+1} + \frac{1}{\rho} y^n),
\end{equation}
where the soft-thresholding operator is defined as
\begin{equation*}
    S_{\lambda}(x) = \text{sign}(x) \max(|x|-\lambda, 0).
\end{equation*}

Update $y$:
\begin{equation}\label{eqn:yupdate}
   y^{n+1} = y^{n} + \rho(D u^{n+1} - z^{n+1}).
\end{equation}

A summary of the proposed ADMM algorithm is shown in Algorithm~\ref{alg:ADMM}.
\begin{algorithm}
\caption{An ADMM algorithm to solve Eqn.~\eqref{eqn:tvregreg3}}\label{alg:ADMM}
\begin{algorithmic}
\State Initialization: choose $\lambda>0$, $\rho>0$, $z^0=0$, $y^0=0$
\State Compuate the Cholesky decomposition of $F = R^T R$
\For{$n$ = 0, 1, $\dots$, Max\_Iter}
\State Update $u^{n+1}$ according to Eqn.~\eqref{eqn:uupdate2}
\State Update $z^{n+1}$ according to Eqn.~\eqref{eqn:zupdate2}
\State Update $y^{n+1}$ according to Eqn.~\eqref{eqn:yupdate}
\If {stopping criteria is met}
    \State Break
\EndIf
\EndFor
\end{algorithmic}
\end{algorithm}

The algorithm steps when the primal and dual residuals~\cite{boyd2011distributed} satisfying a certain stopping criterion, or when the maximal iteration step is reached. The stopping criterion can be specified by two thresholds: absolute tolerance $\epsilon_{abs}$ and relative tolerance $\epsilon_{red}$, see~\cite{boyd2011distributed}. A fixed $\rho$ (say 10) is commonly used. But there are some schemes of varying the penalty parameter to achieve better convergence. In our case, we want to compute a solution path, which requires to solve Eqn.~\eqref{eqn:tvregreg3} with different $\lambda$'s, ordered from large to small. In this case, the Cholesky decomposition of $F$ only needs to be computed once, for the largest $\lambda$. For latter (smaller) $\lambda$'s, one can use the solution of the last $\lambda$ to be the initial guess of the current problem.

In practice, $D$ does not need to be assembed and stored explicitly. Based on the definition of $\|D u\|_1$, see Eqn.~\eqref{eqn:TV2}, we can see that $D$ is a block diagonal matrix and all the computations involved with $D$ can be computed efficiently. For example, $D^T D$ in Eqn.~\eqref{eqn:Fandb} can be computed by
\begin{equation*}
    D^T D = \text{diag}\{w_k^2 D_k^T D_k\}_{k=1}^K.
\end{equation*}
To compute $D y$ for any vector $y$, like in Eqn.~\eqref{eqn:Fandb}, \eqref{eqn:zupdate2} and \eqref{eqn:yupdate}, the computation can be done blockwisely, i.e.,
\begin{equation*}
    (D y)_k = w_k D_k y_k, \qquad 1 \le k \le K.
\end{equation*}

\section{Equivalence of two optimization problems}\label{app:equiv}
Suppose $y^{(0)}(x) = \Expect[Y^{(0)}_n | x]$ is the conditional mean response for the control group, and $\Var^{(0)}(x) = \Var[Y^{(0)}_n | x]$ is the conditional variance. Similarily we define $y^{(1)}(x)$ and $\Var^{(1)}(x)$ for the treatment group. From definition, we know that
\begin{equation}\label{def:tau2}
	y^{(1)}(x) = y^{(0)}(x) + \tau(x).
\end{equation}
We ``pretend'' to fit both $y^{(0)}$ and $y^{(1)}$ by minimizing the least square loss:
\begin{equation}\label{eqn:ls0}
    \min_{y^{(0)}, \tau} \sum_{i=1}^N (Y_n - y^{(W_n)})^2/\Var^{(W_n)}(x).
\end{equation}
After separating the control and treatment groups and merging all measurements at the same covariate vector $x$ together, the LS above is equivalent to 
\begin{equation}\label{eqn:ls1}
    \min_{y^{(0)}, \tau} \sum_x \sum_{w = 0, 1} M^{(w)}(x) (\bar{y}^{(w)}(x) - y^{(w)}(x))^2, 
\end{equation}
where $N^{(w)}(x)$ is the number of measurements with covariate vector $x$ in the $w$-group (0 for control and 1 for treatment), $\bar{y}^{(w)}(x)$ is the corresponding sample mean and $M^{(w)}(x) = N^{(w)}(x)/\Var^{(W_n)}(x)$, i.e.,
\begin{equation*}
	N^{(w)}(x) := \sum_{X_n = x, W_n = w} 1\,, \quad \bar{y}^{(w)}(x) = \frac{\sum_{X_n = x, W_n = w} Y_n}{N^{(w)}(x)}.
\end{equation*}
If we directly solve~\eqref{eqn:ls1}, we gets an estimation of the treatment effect
\begin{equation}
	\hat{\tau}(x) = \bar{y}^{(1)}(x) - \bar{y}^{(0)}(x).
\end{equation}
However, although it is unbiased, the estimation above has a large variance and is difficult to summarize or interpret. Typically, we believe that the treatment effect has some good structures, e.g. sparse effect, piece-wise constant effect and low-order effect. One popular and successful way to enforce these structures in statistics and machine learning is to add an regularization term to the objective function, i.e.
\begin{equation}\label{eqn:master0}
    \min_{y^{(0)}, \tau} \frac{1}{2}\sum_x \sum_{w = 0, 1} M^{(w)}(x) (\bar{y}^{(w)}(x) - y^{(w)}(x))^2 + \lambda \Omega(\tau), 
\end{equation}
where $\Omega(\tau)$ is a regularization of the treatment effect $\tau$ and $\lambda > 0$ is a regularization parameter. 

Since the minimization over $y^{(0)}$ is purely a unconstrained quadratic problem and we can analytically do the computation. After minimizating over $y^{(0)}$, our regularized regression problem~\eqref{eqn:master} becomes an optimization problem only involved with $\tau$:
\begin{equation}
    \min_{\tau} \frac{1}{2}\sum_{x} M_{e}(x) (\hat{\tau}(x) - \tau(x))^2 + \lambda \Omega(\tau),
\end{equation}
where 
\begin{equation}
\begin{split}
    M_{e}(x) & = \frac{M^{(0)}(x)M^{(1)}(x)}{M^{(0)}(x)+M^{(1)}(x)} \\
    & = \frac{N^{(0)}(x)N^{(1)}(x)}{N^{(0)}(x)\Var^{(1)}(x)+N^{(1)}(x)\Var^{(0)}(x)}
\end{split}
\end{equation}
can be viewed as the effective sample size for the noisy measurement $\hat{\tau}(x)$ of the treatment effect at $x$.

We mentioned that we ``pretend'' to fit both $y^{(0)}$ and $y^{(1)}$ because we only add regularity constraints on the treatment effect $\tau$ and fit a model only for $\tau$ ultimately by~\eqref{eqn:master}. When purely looking at Eqn.~\eqref{eqn:master}, we can think that we are applying standard supervised learning methods on noisy measurement $\hat{\tau}(x)$ with mean $\tau(x)$ and covariance $\text{diag}\{1/M_{e}(x)\}$.

The variance of $\Var^{(w)}(x)$ can be directly obtained from historical data, or be estimated at each point $x$ by
\begin{equation}\label{eqn:variance}
	\hat{\Var}^{(w)}(x) = \frac{\sum_{X_n = x, W_n = w} (Y_n - \bar{y}^{(w)}(x))^2}{N^{(w)}(x) - 1}.
\end{equation}
The above estimation~\eqref{eqn:variance} is preferred when we have sufficient measurements in this $(w,x)$-subgroup, i.e., $N^{(w)}(x)$ is large. If we have only limited data, it is preferred to use historical data, or to make assumptions like variance in the control/treatment group is a constant to improve the quality of estimating variance.

\end{document}

%% file: macro.tex
\usepackage{mathtools}
\mathtoolsset{centercolon}  
\usepackage{bm} 
\usepackage{ntheorem}

\newcounter{exc}

\newcounter{quesc}

\newtheorem*{defi-no}[thmc]{Definition}

\newtheorem{problem}[quesc]{Problem}
\newtheorem{ex}[exc]{Example}

\newtheorem{theorem}{Theorem}

\newcommand{\ts}{\textsuperscript}

\renewcommand{\phi}{\varphi}

\providecommand{\mathbbm}{\mathbb} 

\newcommand{\R}{\mathbbm{R}}

\newcommand{\Expect}{\operatorname{\mathbb{E}}}
\newcommand{\Var}{\operatorname{\mathrm{var}}}

\newcommand{\vct}[1]{\bm{#1}}
\newcommand{\mtx}[1]{\bm{#1}}

\renewcommand{\hat}{\widehat}

\DeclareMathOperator*{\argmin}{arg\,min}

\usepackage{xspace}

\newcommand{\iid}{\emph{i.i.d.}\xspace}